\documentclass[aap,MSNbibl,nameyear,dvips]{arximspdf}
\usepackage{algorithm}
\usepackage{accents}

\doi{10.1214/12-AAP909} %kopijuoti is PTS
\volume{23}
\issue{6}
\pubyear{2013}
\firstpage{2500}
\lastpage{2537}

\makeatletter

\newcommand{\mathrmm}{}

\renewcommand{\underline}{\underaccent{\bar}}

\newcommand{\rrVert}{\Vert}
\newcommand{\rrvert}{\vert}
\newcommand{\llVert}{\Vert}
\newcommand{\llvert}{\vert}

\newcommand{\interleave}{\Vert\hspace*{-1.4pt}\vert}

\floatname{algorithm}{Algorithm}

\newproclaim{defn}{Definition}
\newproclaim{rem}{Remark}

\newtheorem{lem}{Lemma}
\newtheorem{prop}{Proposition}
\newtheorem{theorem}{Theorem}
\newtheorem{cor}{Corollary}

\newcommand{\iid}{\stackrel{\mathrm{i.i.d.}}{\sim}}

\makeatother

\begin{document}
\begin{frontmatter}

\title{Stability properties of some particle filters}
\runtitle{Stability properties of some particle filters}

\begin{aug}
\author[A]{\fnms{Nick} \snm{Whiteley}\corref{}\ead[label=e1]{nick.whiteley@bristol.ac.uk}}
\runauthor{N. Whiteley}
\affiliation{University of Bristol}
\address[A]{Department of Mathematics\\
University of Bristol\\
University Walk\\
Bristol\\
BS8 1TW\\
United Kingdom\\
\printead{e1}} %adresu isvedimo komanda gale!
\end{aug}

% HISTORY:
\received{\smonth{10} \syear{2011}}
\revised{\smonth{9} \syear{2012}}

% ABSTRACT
%
\begin{abstract}
Under multiplicative drift and other regularity conditions, it is
established that the asymptotic variance associated with a particle
filter approximation of the prediction filter is bounded uniformly in
time, and the nonasymptotic, relative variance associated with a
particle approximation of the \mbox{normalizing} constant is bounded linearly
in time. The conditions are demonstrated to hold for some hidden Markov
models on noncompact state spaces. The particle stability results are
obtained by proving $v$-norm multiplicative stability and exponential
moment results for the underlying Feynman--Kac formulas.
\end{abstract}

% KEYWORDS

\end{frontmatter}

%s1 #&#
\section{Introduction}\label{intro}

Particle filters have become very popular devices for approximate
solution of nonlinear filtering problems in hidden Markov models
(HMMs) and various aspects of their theoretical properties are now
well understood. However, there are still very few results which establish
some form of stability over time of particle filtering methods on
noncompact spaces, at least without resorting to algorithmic modifications
which involve a random computational expense. The aim of the present
work is to establish theoretical guarantees about some stability properties
of a standard particle filter, under assumptions which are verifiable
for some HMMs with noncompact state spaces.

It is now well known that, under mild conditions, the error associated
with particle approximation of filtering distributions satisfies a
central limit theorem. The first stability property we obtain is a
time-uniform bound on the corresponding asymptotic variance. Making
use of some recent results on functional expansions for particle approximation
measures, the second stability property we obtain is a linear-in-time
bound on the nonasymptotic, relative variance of the particle approximations
of normalizing constants. These two properties are established by
first proving some multiplicative stability and exponential moment
results for the Feynman--Kac formulas underlying the particle filter.
The adopted approach involves Lyapunov function, multiplicative stability
ideas in a weighted $\infty$-norm setting, which allows treatment
of a noncompact state space. We thus obtain stability results which
hold under weaker assumptions than those existing in the literature.
The main restriction is that our assumptions are typically satisfied
under some constraints on the observation component of the HMM and/or
the observation sequence driving the filter. On the other hand, subject
to these constraints, our stability results hold uniformly over observation
records and without any stochasticity necessarily present in the observation
process.

The rest of this paper is structured as follows. Section \ref{secSetting}
briefly introduces filtering in HMMs and particle filters and comments
on some existing stability results. Section \ref{subApplication-of-some}
gives some applications of the main particle stability results to
classes of hidden Markov models. The hope is that Sections \ref{secSetting}
and \ref{subApplication-of-some} can be read without the reader
necessarily delving into the main results of Section \ref
{sec-stability-of-Feynman-Kac}
or the corresponding proofs and auxiliary results of Section \ref{secProofs},
which are obtained in the more abstract setting of interacting particle
approximations of Feynman--Kac formulas.

%s2 #&#
\section{Setting}\label{secSetting}

%s2.1 #&#
\subsection{Hidden Markov models and filtering}

A hidden Markov model is a bi-variate, discrete-time Markov chain
$ ( (X_{n},Y_{n} );n\geq0 )$ where the signal process
$ (X_{n} )$ is also a Markov chain and each observation
$Y_{n}$ is conditionally independent of the rest of the bi-variate
process given $X_{n}$. Each $X_{n}$ is valued in a state-space~$\mathsf{X}$,
and each $Y_{n}$ is valued in the observation space $\mathsf{Y}$.
The present work focuses on the case where $\mathsf{X}$ is noncompact,
and we are typically interested in the case that $\mathsf{X}$ is
some subset of $\mathbb{R}^{d}$. In any case, throughout the following
we assume that $\mathsf{X}$ and $\mathsf{Y}$ are Polish spaces endowed
with their respective Borel $\sigma$-algebras, $\mathcal{B} (\mathsf
{X} )$~and $\mathcal{B} (\mathsf{Y} )$. Our main stability results,
presented in Section \ref{sec-stability-of-Feynman-Kac}, are in
the setting of Feynman--Kac formulas which can be considered as underlying
the filtering problem of interest. In that section, more precise definitions
are given. In the present section, we consider the HMM directly.

Let $\mu$ be a probability distribution on $\mathsf{X}$, let $f$
be a Markov kernel acting from $\mathsf{X}$ to itself and let $g$
be a Markov kernel acting from $\mathsf{X}$ to $\mathsf{Y}$, with
$g(x,\cdot)$ admitting density, similarly denoted by $g (x,y )$,
with respect to some dominating $\sigma$-finite measure. We will
assume that $g (x,y )>0$ and, for now, that $\sup_{x,y}g(x,y)<\infty$.
Loosely speaking, the task of filtering is to compute some conditional
distributions of the $ (X_{n} )$ process given the observations
$ (Y_{n} )$, under an assumed model,
%
%e1 #&#
\begin{eqnarray}\label{eqHMM}
(X_{0},Y_{0} )&\sim&\mu(dx_{0} )g
(x_{0},dy_{0} ),
\nonumber\\[-8pt]\\[-8pt]
(X_{n},Y_{n} )| \{ X_{n-1}=x_{n-1}
\} &\sim& f (x_{n-1},dx_{n} )g (x_{n},dy_{n}
),\qquad n\geq1.\nonumber
\end{eqnarray}
For a realization of observations $ (y_{0},y_{1},\ldots)$,
we may take as a recursive definition of the (one-step-ahead) \textit{prediction
filters}, the sequence of distributions $ (\pi_{n};n\geq0 )$
following
%
%e2 #&#
\begin{eqnarray}\label{eqfilteringrecursion}
\pi_{0} (dx_{0} )&:=&\mu(dx_{0} ),
\nonumber\\[-8pt]\\[-8pt]
\pi_{n} (dx_{n} )&:=&\frac{\int_{\mathsf{X}}\pi_{n-1}
(dx_{n-1} )g (x_{n-1},y_{n-1} )f(x_{n-1},dx_{n})}{\int_{\mathsf{X}}\pi
_{n-1} (dx_{n-1} )g (x_{n-1},y_{n-1}
)},\qquad n\geq1.\nonumber
\end{eqnarray}
We also define the sequence $ (Z_{n};n\geq0 )$ by
%
%e3 #&#
\begin{equation}\label{eqZrecusion}
Z_{0}:=1,\qquad Z_{n}:=Z_{n-1}\int
_{\mathsf{X}}\pi_{n-1} (dx_{n-1} )g
(x_{n-1},y_{n-1} ),\qquad n\geq1.
\end{equation}
Note that the dependence of $\pi_{n}$ and $Z_{n}$ on $y_{0:n-1}=
(y_{0},\ldots, y_{n-1} )$
is suppressed from the notation. Unless stated otherwise, whenever
$ (\pi_{n} )$ or $ (Z_{n} )$ appear below it should
be understood that they depend on an arbitrary but fixed and deterministic
$\mathsf{Y}$-valued sequence $ (y_{0},y_{1},\ldots)$.
The same applies for the particle approximations introduced in Section
\ref{subParticle-filtering}. The set of observation sequences for
which our particle variance results hold is made precise and discussed
in Section \ref{subApplication-of-some}.

Under model (\ref{eqHMM}), $\pi_{n}$ is the conditional distribution
of $X_{n}$ given $ \{ Y_{0:n-1}=y_{0:n-1} \} $; and $Z_{n}$
is the joint density of $Y_{0:n-1}$ evaluated at $y_{0:n-1}$. The
convention of working with the one-step-ahead quantities is mostly
for simplicity of presentation in the following.

In applications there typically will be some degree of model mis-specification;
perhaps the data generating process $ (X_{n},Y_{n} )$ is
not distributed according to (\ref{eqHMM}) with this particular
$\mu$, $f$ and $g$, or perhaps $ (Y_{n} )$ are not the
observations from an HMM at all [for ease of presentation we purposefully
avoid giving a name to a ``true'' distribution for $ (Y_{n} )$].
Nevertheless, as $ (y_{0},y_{1},\ldots)$ arrive our aim
is to compute, or well-approximate $ (\pi_{n} )$ and $
(Z_{n} )$
as per (\ref{eqfilteringrecursion})--(\ref{eqZrecusion}) with
some $\mu$, $f$ and $g$ of our choosing.

HMMs are simple and yet flexible models which have found countless
applications. However, under choices of $\mu$, $f$ and $g$ which
are desirable in many practical situations, $ (\pi_{n} )$
and $ (Z_{n} )$ are not available in closed form.

%s2.2 #&#
\subsection{Particle filtering}\label{subParticle-filtering}

Particle filters [\citet{gordon1993novel}] are a class of stochastic
algorithms which yield approximations of $ (\pi_{n} )$ and
$ (Z_{n} )$ using a population of $N$ samples which interact
over time. These approximations will be denoted by $ (\pi
_{n}^{N} )$
and $ (Z_{n}^{N} )$. Algorithm \ref{algpf} is perhaps
the most simple generic particle filtering scheme (a more precise
probabilistic definition is considered in Section \ref
{sec-stability-of-Feynman-Kac}).
At time $n\geq1$, the sampling step performs a selection--mutation
operation and is equivalent to choosing, with replacement, $N$ individuals
from the population on the basis of their fitness, proportional to
$g (\cdot,y_{n-1} )$, followed by them each mutating in
a conditionally independent manner according to $f$.

\begin{algorithm}%[H]
\caption{}\label{algpf}
For $n=0$,

\qquad Sample $ (\xi_{0}^{i} )_{i=1}^{N}\iid\mu$,\vspace*{2pt}

\qquad Report $\pi_{0}^{N}={ \frac{1}{N}\sum_{i=1}^{N}\delta_{\xi
_{0}^{i}}}$, $Z_{0}^{N}=1$.

For $n\geq1$,

\qquad Report $Z_{n}^{N}=Z_{n-1}^{N}{\frac{1}{N}\sum
_{j=1}^{N}g (\xi_{n-1}^{j},y_{n-1} )}$,

\qquad Sample $ (\xi_{n}^{i} )_{i=1}^{N}| (\xi
_{n-1}^{i} )_{i=1}^{N} \iid\frac{\sum_{j=1}^{N}g (\xi
_{n-1}^{j},y_{n-1} )f (\xi_{n-1}^{j},\cdot)}{\sum_{j=1}^{N}g (\xi
_{n-1}^{j},y_{n-1} )}$,

\qquad Report $\pi_{n}^{N}={ \frac{1}{N}\sum_{i=1}^{N}\delta_{\xi_{n}^{i}}}$.
\end{algorithm}

A large number of variations and extensions of this algorithm have
been developed. A full survey is well beyond the scope of the present
work, but a few comments are called for. Firstly, Algorithm \ref{algpf}
implicitly uses ``multinomial resampling'' at every time step.
It would be interesting to investigate similar results to those presented
here for other resampling schemes, for example, via the analyses of
\citet{smctheC04,smctheDMJDJ11}. Second, Algorithm~\ref{algpf}
involves mutation at every time step according to the Markov kernel
$f$. Again, various alternative schemes have been devised. Mutation
according to $f$ is not an essential characteristic of the main results
of Section \ref{sec-stability-of-Feynman-Kac}, and it is only for
simplicity that the results of Section \ref{subApplication-of-some}
are presented in this context. Third, the results presented here
are likely to be relevant to related classes of sequential Monte Carlo
methods, for example, the smoothing algorithms treated by \citet{del2010backward}
and \citet{Douc2010smoothing}.

%s2.3 #&#
\subsection{Existing stability results for particle filters}

One of the first and most influential works on stability of particle
filters is that of \citet{smctheDMG01} who established time-uniform
convergence properties of the particle approximations. They required
uniform upper and lower bounds on $g$ and stability of the corresponding
exact filter, in turn derived using quite strong assumptions on $f$
involving simultaneous, uniform minorization and majorization, which
are rarely satisfied then $\mathsf{X}$ is noncompact. Similar mixing
assumptions have been employed in \citet
{smctheLGO04,smctheC04,smctheK05,smctheCdMG11}
in order to establish (resp.) uniform convergence of particle
filtering approximations; a time-uniform bound on the asymptotic variance;
and linear-in-time bounds on the nonasymptotic variance of the normalizing
constant estimate. All also consider variants of the standard particle filter
in Algorithm \ref{algpf}.

\citet{legland2003robustification} developed truncation ideas in
order to achieve uniform particle approximations without mixing assumptions,
but with random computational cost and/or proposals restricted to
compact sets. A further development was made by \citet{smctheOR05},
allowing treatment of some nonergodic signals via a particle filter
incorporating an accept/reject step. Truncation ideas have also been
used in \citet{smctheHC08} in order to obtain uniform convergence
of particle filter approximations for HMMs on noncompact state-spaces
with quite specific structures (including $\mathsf{X}$ and $\mathsf{Y}$
being of the same dimension). \citet{smcthevH09} has established
uniform convergence of time-averaged filters under tightness assumptions
on noncompact spaces. \citet{del2001interacting} proved tightness
of the sequence of asymptotic variances (as a function of random observations)
in the linear-Gaussian case. \citet{favetto2009asymptotic} has proved
tightness of the same for a class of HMMs, but subject to a mixing
assumption on~$f$.

It is stressed that: (1) a time-uniform bound on the asymptotic variance
for $\pi_{n}^{N}$, and (2) a linear-in-time bound on the relative variance
for $Z_{n}^{N}$, as pursued here, are different properties from
the time-uniform convergence results proved in most of the above.
The existing works featuring the most similar type of results to those
considered here are \citet{smctheC04,smctheK05,favetto2009asymptotic}
and \citet{smctheCdMG11}, all of which rely on strong mixing assumptions,
at least on $f$, which we do not invoke.

The overall approach used in the present work to express Feynman--Kac
formulas and associated functionals is the semigroup formulation of
\citet{smctheoryDm04}, but the stability ideas are different and
are based around a weighted $\infty$-norm function space setting.
In Theorem \ref{thmMET}, the decomposition idea of \citet{filtertheKV08}
and some technical approaches from \citet{filtertheDFMP09} are
employed.

For completeness we also mention the following. \citet{whiteley2011a}
considered stability properties of a related class of sequential Monte
Carlo methods which are not used for filtering and operate in a different
structural regime, where the number of distributions involved may
be considered a parameter of the algorithm. \citet{whiteley2011}
considered relative variance for $Z_{n}^{N}$ in the context of time-homogeneous
Feynman--Kac models (obtained in the present setting by setting all
$y_{0},y_{1},\ldots$ to a constant), appealing to spectral properties
of the integral kernel involved. There is nothing explicitly spectral
about the present work, but there are some related structural ideas
involved; see Section \ref{sec-stability-of-Feynman-Kac}. For example,
Theorem \ref{thmMET} is expressed in such a way that it may be viewed
as an nonhomogeneous analogue of the $v$-norm multiplicative ergodicity
results of \citet{mctheKM05}, in the context of positive operators.
The assumptions in the present work also allow the treatment of time-homogeneous
Feynman--Kac models, and in that setting are actually stronger than
the assumptions of \citet{whiteley2011} [because in assumption
\ref{hypminoronly}--\ref{hypminorandmaj}
of Section \ref{subUniform--controls} here, we require a simultaneous
local minorization/majorization condition], but on the other hand
the approach of \citet{whiteley2011} is specific to the time-homogeneous
setting.

%s3 #&#
\section{Summary and application of some results}\label{subApplication-of-some}

In this section, the results of Section \ref{sec-stability-of-Feynman-Kac}
are summarized and applied to some specific hidden Markov models and
the particle filter of Algorithm \ref{algpf}. To this end we consider
the following assumptions on $\mu$, $f$ and $g$ which serve as
an intermediate layer of abstraction and which together imply that
assumptions \ref{hypdrift}--\ref{hypGbounded} of Section
\ref{sec-stability-of-Feynman-Kac} are satisfied. Discussion of
the latter assumptions and their relation to the existing literature
is given in Section \ref{subComments-on-the}.

Consider the following:
\begin{itemize}
\item$\mathsf{Y}_{\star}\subseteq\mathsf{Y}$ is measurable,
and the quantities in the below conditions may depend on~$\mathsf{Y}_{\star}$.
\item There exists $V\dvtx \mathsf{X}\rightarrow[1,\infty)$ unbounded,
$\underline{d}\in[1,\infty)$ and $\delta>0$ with the
following properties. For each $d\in[\underline{d},\infty)$,
%
%e4 #&#
\begin{equation}\label{eqaccesiblefront}
g(x,y)\int_{C_{d}}f\bigl(x,dx'\bigr)>0\qquad \forall x
\in\mathsf{X}, y\in\mathsf{Y}_{\star},
\end{equation}
where $C_{d}:= \{ x\dvtx V(x)\leq d \} $, and there exists
$b_{d}<\infty$
such that
%
%e5 #&#
\begin{eqnarray}\label{eqdriftfront}
& & \sup_{y\in\mathsf{Y}_{\star}}g(x,y)\int_{\mathsf
{X}}f
\bigl(x,dx'\bigr)\exp\bigl[V\bigl(x'\bigr) \bigr]
\nonumber\\[-8pt]\\[-8pt]
& &\qquad \leq\exp\bigl[V(x) (1-\delta)+b_{d}\mathbb{I}_{C_{d}} (x
) \bigr]\qquad \forall x\in\mathsf{X},\nonumber
\end{eqnarray}
and there exists a probability measure $\nu_{d}$ and $0<\varepsilon
_{d}^{-}\leq\varepsilon_{d}^{+}<\infty$
such that
%
%e6 #&#
\begin{eqnarray}\label{eqminormajorfront}
& & \varepsilon_{d}^{-}\nu_{d} \bigl(dx'
\bigr)\mathbb{I} {}_{C_{d}} \bigl(x^{\prime} \bigr)
\nonumber
\\
& &\qquad \leq g (x,y )f \bigl(x,dx^{\prime} \bigr)\mathbb{I} {}_{C_{d}}
\bigl(x^{\prime} \bigr)
\\
& &\qquad \leq\varepsilon_{d}^{+}\nu_{d}
\bigl(dx' \bigr)\mathbb{I} {}_{C_{d}} \bigl(x^{\prime}
\bigr)\qquad \forall x\in C_{d}, y\in\mathsf{Y}_{\star}\nonumber
\end{eqnarray}
with $\nu_{d} (C_{r} )>0$ for all $r\in[\underline
{d},d ]$.
\item$\int\exp[V(x) ]\mu(dx )<\infty$.
\item Although not required for all results of Section \ref
{sec-stability-of-Feynman-Kac},
in the present section it is also assumed that
%
%e7 #&#
\begin{equation}\label{eqgboundedfront}
\sup_{ (x,y )\in\mathsf{X}\times\mathsf{Y}_{\star
}}g(x,y)<\infty.
\end{equation}
\end{itemize}
The condition of (\ref{eqdriftfront}) is a multiplicative drift
condition. Similar conditions have been used in the study of stability
of exact filters [\citet{filtertheDFMP09}] and can hold when
$\mathsf{Y}_{\star}=\mathsf{Y}$ is noncompact. It may be the case that
$f$ alone satisfies such a multiplicative condition (see Section
\ref{subergsignalmodel} below), in which case (\ref{eqdriftfront}) can
be satisfied when $\sup_{y\in\mathsf{Y}}g(x,y)$ is not bounded above in
$x$. When (\ref{eqgboundedfront}) holds, then (\ref{eqdriftfront}) can
hold even when $f$ is not ergodic, but it is then typically required
that $\mathsf{Y}_{\star}\subset\mathsf{Y}$ is compact; see Section
\ref{subA-class-ofnonergodic}. The conditions of (\ref
{eqminormajorfront}) and (\ref{eqgboundedfront}) together imply that
for all $d\in [\underline{d},\infty)$,
\[
\sup_{y\in\mathsf{Y}_{\star}}\sup_{ (x,x^{\prime}
)\in C_{d}\times C_{d}}\frac{g (x,y )}{g (x^{\prime
},y )}<
\infty,
\]
which can, loosely, be interpreted as a constraint on the amount of
information which any single observation in $\mathsf{Y}_{\star}$
can provide about the hidden state in each~$C_{d}$. For the example
of Section \ref{subDiscrete-valued-observations} we are able to
satisfy the assumptions when $\mathsf{Y}_{\star}=\mathsf{Y}$
is compact. For noncompact $\mathsf{Y}$ in the examples below, we
resort to taking $\mathsf{Y}_{\star}$ compact.

Under the above assumptions, the main conclusions of Propositions
\ref{propvarianceunibound} and \ref{propvariance}, Section \ref
{subVariance-bounds},
may be summarized as follows.\vspace*{9pt}

\textit{Uniformly bounded variance in the CLT for $\pi_{n}^{N}$.\quad}
It is known [e.g., \citet{smctheoryDm04}, Section 9.4.2] that under
(\ref{eqgboundedfront}), for any $\varphi\dvtx \mathsf{X}\rightarrow
\mathbb{R}$
bounded, measurable, $n\geq1$ and any $y_{0:n}\in\mathsf{Y}_{\star}^{n+1}$,
\[
\sqrt{N}\int_{\mathsf{X}} \bigl[\pi_{n}^{N}(dx)-
\pi_{n}(dx) \bigr]\varphi(x)\longrightarrow\mathcal{N} \bigl(0,
\sigma_{n}^{2} (y_{0:n} ) \bigr)
\]
in distribution as $N\rightarrow\infty$. Under the conditions of
(\ref{eqaccesiblefront})--(\ref{eqgboundedfront}), Proposition
\ref{propvarianceunibound} may be applied to establish there exists
$c_{\mu}<\infty$ depending on $\mathsf{Y}_{\star}$, such that for
all such $\varphi$ and $n\geq0$
%
%e8 #&#
\begin{equation}\label{eqCLTfront}
\sigma_{n}^{2} (y_{0:n} )\leq
\operatorname{Var}_{\pi_{n}} (\varphi)+\llVert\varphi\rrVert
^{2}c_{\mu}\qquad \forall y_{0:n}\in\mathsf{Y}_{\star}^{n+1}
\end{equation}
with \mbox{$\llVert\cdot\rrVert$} the sup norm. Discussion of a
CLT for other classes of $\varphi$ is given in Section \ref
{subAsymptotic-variance-for}.\vspace*{9pt}

\textit{Linearly bounded relative variance for $Z_{n}^{N}$.\quad}
Under the conditions of (\ref{eqaccesiblefront})--(\ref{eqgboundedfront}),
Proposition \ref{propvariance} may be applied to establish that
there exists $c_{\mu}'<\infty$ depending on $\mathsf{Y}_{\star}$
such that for all $n\geq0$,
%
%e9 #&#
\begin{eqnarray}\label{eqvarfront}
N>c_{\mu}' (n+1 ) \quad\Longrightarrow\quad\mathbb{E}_{\mu}
\biggl[ \biggl(\frac{Z_{n}^{N}}{Z_{n}}-1 \biggr)^{2} \biggr]\leq
c_{\mu}'\frac
{4}{N} (n+1 )\nonumber\\[-8pt]\\[-8pt]
&&\eqntext{\forall
y_{0:n}\in\mathsf{Y}_{\star
}^{n+1},}
\end{eqnarray}
where $\mathbb{E}_{\mu}$ is expectation with respect to the law of
the $N$-particle filtering algorithm initialized using $\mu$.

%s3.1 #&#
\subsection{A class of ergodic signal models}\label{subergsignalmodel}

The following class of signal models has been considered by \citet
{filtertheKV08}
and \citet{filtertheDFMP09} in the context of stability of exact
filters (i.e., without particle approximation). We have $\mathsf
{X}=\mathbb{R}^{d_{x}}$
for some $d_{x}\geq$1. The transition kernel $f$ corresponds to
the signal model
%
%e10 #&#
\begin{equation}\label{eqKVmodel}
X_{n+1}=X_{n}+B(X_{n})+\sigma(X_{n})W_{n},\qquad (W_{n};n\geq1
)\iid\mathcal{N} (0,I_{d_{x}})
\end{equation}
with:
\begin{itemize}
\item$B$ is a $d_{x}$-dimensional vector function, locally bounded and
%
%e11 #&#
\begin{equation}\label{eqKVmodelb}
\lim_{r\rightarrow\infty}\sup_{\llvert x\rrvert\geq r}\bigl\llvert x+B(x)
\bigr\rrvert-\llvert x\rrvert=-\infty;
\end{equation}

\item$\sigma$ is a $d_{x}\times d_{x}$ matrix function, and has the so-called
nondegenerate noise variance property
%
%e12 #&#
\begin{eqnarray}\label{eqKVmodelsigma}
0&<&\inf_{x\in\mathbb{R}^{d_{x}}}\inf_{\lambda\in\mathbb
{R}^{d_{x}},\llvert\lambda\rrvert=1}
\lambda^{T}\sigma(x)\sigma^{T}(x)\lambda
\nonumber\\[-8pt]\\[-8pt]
&\leq&\sup_{x\in\mathbb{R}^{d_{x}}}\sup_{\lambda\in\mathbb
{R}^{d_{x}},\llvert\lambda\rrvert=1}
\lambda^{T}\sigma(x)\sigma^{T}(x)\lambda< \infty.\nonumber
\end{eqnarray}
\end{itemize}

As per Lemma \ref{lemARsignalmodel} in Section \ref{secProofs}, $f$ in this
case itself satisfies a multiplicative drift condition with $v(x):=\exp
(1+c\llvert x\rrvert )$
for $c$ a positive constant. An example of a possible signal model
with non-Gaussian transition probability and $f$ itself satisfying
a multiplicative drift condition is the discretely sampled Cox--Ingersoll--Ross
process; see \citet{whiteley2011}.

We now discuss some observation models which may be combined with
the signal model above.

%s3.1.1 #&#
\subsubsection{Discrete-valued observations}\label{subDiscrete-valued-observations}

With $\mathsf{Y}= \{ 0,1 \} ^{d_{x}}$, consider the multivariate
binary observation model
\[
\bigl(Y_{n}^{1},\ldots,Y_{n}^{d_{x}}
\bigr)| \{ X_{n}=x_{n} \} \sim\operatorname{\mathcal{B}e} \bigl(p
\bigl(x_{n}^{1} \bigr) \bigr)\otimes\cdots\otimes
\operatorname{\mathcal{B}e} \bigl(p \bigl(x_{n}^{d_{x}} \bigr) \bigr),
\]
where $\mathcal{B}\mathrm{e}$ denotes the Bernoulli distribution, $p(x):=1/
(1+e^{-x} )$
and $Y_{n}= (Y_{n}^{1},\ldots,Y_{n}^{d_{x}} )$, $x_{n}=
(x_{n}^{1},\ldots,x_{n}^{d_{x}} )$.
This corresponds to
\[
g(x,y)=\prod_{j=1}^{d_{x}}p
\bigl(x^{j} \bigr)^{\mathbb{I}
[y^{j}=1 ]} \bigl(1-p \bigl(x^{j} \bigr)
\bigr)^{\mathbb{I}
[y^{j}=0 ]}.
\]
Clearly $\sup_{x,y}g(x,y)=1$ and for any compact $C\subset\mathbb{R}^{d_{x}}$,
$\inf_{x\in C}\inf_{y\in\mathsf{Y}}g(x,y)>0$. Combined with Lemma
\ref{lemARsignalmodel}, this establishes that the assumptions
of equations (\ref{eqaccesiblefront}), (\ref{eqdriftfront})
and (\ref{eqminormajorfront}) are satisfied when this observation
model is combined with the signal model of equations (\ref
{eqKVmodel})--(\ref{eqKVmodelsigma}).

%s3.1.2 #&#
\subsubsection{Uninformative observations in $\mathbb{R}^{d}$}

With $\mathsf{Y}=\mathbb{R}^{d_{y}}$, $d_{y}\geq1$, consider the
observation model
\[
Y_{n}=H (X_{n} )+\zeta_{n},\qquad (
\zeta_{n};n\geq1 )\iid\mathcal{N} (0,I_{d_{y}} )
\]
with $H$ a bounded, vector-function. That the disturbance terms are
standard normal here is only for simplicity of presentation. Obviously
we have
\[
g(x,y)=\frac{1}{ (2\pi)^{d_{y}/2}}\exp\biggl(-\frac{1}{2} \bigl[y-H(x)
\bigr]^{T} \bigl[y-H(x) \bigr] \biggr)
\]
so that $\sup_{(x,y)\in(\mathsf{X},\mathsf{Y} )}g(x,y)=
(2\pi)^{-d_{y}/2}$.
In this case the observations may be considered uninformative as for
each $y$, $\inf_{x\in\mathsf{X}}g(x,y)>0$. In light of Lemma \ref
{lemARsignalmodel},
standard calculations show that this observation model combined with
$f$ of (\ref{eqKVmodel})--(\ref{eqKVmodelsigma}) satisfies
the drift condition of (\ref{eqdriftfront}) with $\mathsf{Y}_{\star
}=\mathsf{Y}$
and $\underline{d}$ chosen large enough. However, when we attempt
to verify (\ref{eqminormajorfront}) [via (\ref{eqKVmodlefmajmin})
in Lemma \ref{lemARsignalmodel}] by incorporating $g(x,y)$, the
minorization part of (\ref{eqminormajorfront}) is not satisfied
with $\mathsf{Y}_{\star}=\mathsf{Y}$, due to the requirement of uniformity
in $y$. We may satisfy (\ref{eqminormajorfront}) by taking $\mathsf
{Y}_{\star}\subset\mathsf{Y}$
a compact set, and the constants involved will then depend on $\mathsf
{Y}_{\star}$.

%s3.1.3 #&#
\subsubsection{Stochastic volatility observations}

With $\mathsf{Y}=\mathbb{R}$ and $d_{x}=1$, consider the stochastic
volatility observation model [considered in \citet{filtertheDFMP09},
Section~4.3],
\[
Y_{n}=\beta\exp(X_{n}/2 )\varepsilon_{n},\qquad (
\varepsilon_{n};n\geq0 )\iid\mathcal{N} (0,I_{d_{x}} ),
\]
where $\beta>0$ is a fixed parameter of the model. The corresponding
likelihood is
\[
g(x,y)=\frac{1}{ (2\pi)^{1/2}\beta}\exp\bigl[-y^{2}\exp(-x )/\bigl(2
\beta^{2}\bigr)-x/2 \bigr],
\]
which is not uniformly upper-bounded on $\mathsf{X}\times\mathsf{Y}$.
But, as noted in Douc et al. [(\citeyear{filtertheDFMP09}), Section
4.3], $\sup_{x\in \mathsf{X}}g(x,y)\leq(2\pi e )^{-1/2}\llvert
y\rrvert^{-1}$. For $0<\underline{y}<\bar{y}<\infty$, take
$\mathsf{Y}_{\star}:= [-\bar{y},-\underline{y}
]\cup[\underline{y},\bar{y} ]$. Then (\ref{eqgboundedfront}) is
satisfied, and using Lemma \ref {lemARsignalmodel}, the drift condition
of (\ref{eqdriftfront}) and the upper bound of
(\ref{eqminormajorfront}) are satisfied with $\underline{d}$ large
enough. The lower bound of (\ref{eqminormajorfront}) is also satisfied
because for $d<\infty$, $\inf_{ (x,y )\in
C_{d}\times\mathsf{Y}_{\star}}g(x,y)>0$.

%s3.2 #&#
\subsection{A class of possibly nonergodic signal models}\label{subA-class-ofnonergodic}

We now consider a class of signal model which includes some nonergodic
$f$ and point out how characteristics of the observation model can
be used to satisfy the drift condition (\ref{eqdriftfront}).

Take $\mathsf{X}=\mathbb{R}^{d_{x}}$ for some $d_{x}\geq1$, and consider
the signal model
%
%e13 #&#
\begin{equation}\label{eqKVmodel-1}
X_{n+1}=B(X_{n})+W_{n},\qquad (W_{n};n\geq0
)\iid\mathcal{N} (0,I_{d_{x}} )
\end{equation}
with $B$ is a $d_{x}$-dimensional vector function, locally bounded.
That the disturbance terms $ (W_{n} )$ are standard normal\vadjust{\goodbreak}
is only for simplicity of presentation; one can draw analogous conclusions
under conditions such as (\ref{eqKVmodelsigma}), but we focus
here on the interplay between $V$, $\mathsf{Y}_{\star}$, $B$ and
$g$. For some $\delta_{0}>1$, take $V(x):=\frac{x^{T}x}{2 (1+\delta
_{0} )}+1$.

Assuming that $\mathsf{Y}_{\star}$, $B$ and $g$ are such that,
for some $\delta_{1}\in(0,1)$,
%
%e14 #&#
\begin{equation}\label{eqexamplesBH}
\lim_{r\rightarrow\infty}\sup_{\llvert x\rrvert\geq r}\sup
_{y\in\mathsf
{Y}_{\star}}-\frac{(1-\delta_{1})x^{T}x}{2 (1+\delta_{0}
)}+\frac{1}{2\delta_{0}}B(x)^{T}B(x)+
\log g(x,y)<0.
\end{equation}
Standard manipulations then establish that the drift condition of
(\ref{eqdriftfront}) is satisfied with $\delta<\delta_{1}$ and
$\underline{d}$ large enough. For the condition of (\ref{eqdriftfront}),
again with $\underline{d}$ large enough we can take $\nu_{d}$ the
normalized restriction of Lebesgue measure to $C_{d}$ if it is the
case that
%
%e15 #&#
\begin{equation}\label{eqexamplegbelow}
\inf_{ (x,y )\in C_{d}\times\mathsf{Y}_{\star}}g(x,y)>0.
\end{equation}

Conditions (\ref{eqexamplesBH}) and (\ref{eqexamplegbelow})
are satisfied, for example, when:
\begin{itemize}
\item the signal model is a random walk, $B(x):=x$;
\item$\mathsf{Y}=\mathbb{R}^{d_{y}}$,
\[
Y_{n}=H(x)+\sigma_{y}\zeta_{n},\qquad (
\zeta_{n};n\geq0 )\iid\mathcal{N} (0,I_{d_{y}} )
\]
with $\sigma_{y}>0$, so that
\[
g(x,y)=\frac{1}{ (2\pi)^{d_{y}/2}\sigma_{y}^{d_{y}}}\exp\biggl(-\frac
{1}{2\sigma_{y}^{2}} \bigl[y-H(x)
\bigr]^{T} \bigl[y-H(x) \bigr] \biggr);
\]

\item$\mathsf{Y}_{\star}$ is compact;
\item$H$ is locally bounded and such that
\end{itemize}
\begin{eqnarray*}
&&\lim_{r\rightarrow\infty}\sup_{\llvert x\rrvert\geq r}\biggl[
\frac
{x^{T}x}{2}\frac{ (1+\delta_{1} )}{\delta_{0}(1+\delta_{0})}
\\
&&\hspace*{24.5pt}\qquad{}+\frac{ (\sup_{y\in\mathsf{Y}_{\star}}\llvert y\rrvert )}{\sigma
_{y}^{2}} \Bigl(\sup_{|\lambda|=1}\lambda^{T}H(x)
\Bigr)-\frac
{H(x)^{T}H(x)}{2\sigma_{y}^{2}}\biggr]<0.
\end{eqnarray*}
Here we observe a trade-off in terms of $\delta_{0}$
(which defines $V$), the constant $\delta_{1}$ ($\delta<\delta_{1}$
appears in the drift condition), the observation noise variance $\sigma_{y}^{2}$
and the growth of $H(x)$.

%s4 #&#
\section{$\mathcal{L}_{v}$-stability of Feynman--Kac formulas and particle
approximations}\label{sec-stability-of-Feynman-Kac}

%s4.1 #&#
\subsection{Definitions and assumptions}\label{subFKsetup}

As per the \hyperref[intro]{Introduction}, let the Polish state space $\mathsf{X}$
be noncompact and endowed with its Borel $\sigma$-algebra $\mathcal
{B}(\mathsf{X})$
(the observation space $\mathsf{Y}$ will\vadjust{\goodbreak} not feature explicitly in
the following Feynman--Kac formulation; see Remark \ref{re2} below). For a
weighting function $v\dvtx \mathsf{X}\rightarrow[1,\infty)$, and $\varphi$
a measurable, real-valued function on $\mathsf{X}$, define the norm
$\llVert\varphi\rrVert_{v}:=\sup_{x\in\mathsf{X}}\llvert\varphi
(x)\rrvert/v(x)$
and let $\mathcal{L}_{v}:= \{ \varphi\dvtx \mathsf{X}\rightarrow\mathbb
{R};\llVert\varphi\rrVert_{v}<\infty\} $
be the corresponding Banach space. Throughout, when dealing with weighting
functions we employ an lower/upper-case convention for exponentiation
and write interchangeably $v\equiv e^{V}$.

For $K$ a kernel on $\mathsf{X}\times\mathcal{B} (\mathsf{X} )$,
a function $\varphi$ and a measure $\eta$ denote $\eta(\varphi):=\int
\varphi(x)\eta(dx)$,
$K\varphi(x)=K (\varphi)(x):=\int K(x,dy)\varphi(y)$ and
$\eta K(\cdot):=\int\eta(dx)\*K(x,\cdot)$. Let $\mathcal{P}$ be the
collection of probability measures on $ (\mathsf{X},\mathcal
{B}(\mathsf{X}) )$,
and for a given weighting function $v\dvtx \mathsf{X}\rightarrow[1,\infty)$,
let $\mathcal{P}_{v}$ denote the subset of such measures $\eta$
such that $\eta(v)<\infty$.

The induced operator norm of a linear operator $K$ acting $\mathcal
{L}_{v}\rightarrow\mathcal{L}_{v}$
is
\[
\interleave K\interleave_{v}:=\sup\biggl\{ \frac{\llVert K\varphi
\rrVert_{v}}{\llVert\varphi\rrVert_{v}};
\varphi\in\mathcal{L}_{v},\llVert\varphi\rrVert_{v}
\neq0 \biggr\} =\sup\bigl\{ \llVert K\varphi\rrVert_{v};\varphi\in
\mathcal{L}_{v},\llvert\varphi\rrvert\leq v \bigr\}.
\]
The corresponding $v$-norm on signed measures is $\llVert
\eta\rrVert_{v}:=\sup_{\llvert\varphi\rrvert\leq v}\llvert\eta
(v )\rrvert$.
For any $n\geq1$ and $1\leq s\leq(n+1 )$, define $\mathcal
{I}_{n,s}:= \{ (i_{1},\ldots,i_{s})\in\mathbb{N}_{0}^{s};0\leq
i_{1}<\cdots<i_{s}\leq n \} $.

Let $\mu\in\mathcal{P}$ be an initial distribution, and for each $n\in
\mathbb{N}$
let $ (M_{n};n\geq1 )$ be a collection of Markov kernels,
each kernel acting $\mathsf{X}\times\mathcal{B}(\mathsf{X})\rightarrow[0,1]$.
Let $ (G_{n};n\geq0 )$ be a collection of $\mathcal{B}(\mathsf
{X})$-measurable,
real-valued, strictly positive functions on~$\mathsf{X}$.

Next let $ (Q_{n};n\geq1 )$ be the collection of integral
kernels defined by
\[
Q_{n}\bigl(x,dx'\bigr):=G_{n-1}(x)M_{n}
\bigl(x,dx'\bigr).
\]
For $1\leq p\leq n$, let $Q_{p,n}$ be the semigroup defined by
%
%e16 #&#
\begin{equation}\label{eqQsemigroup}
Q_{p,n} := Q_{p+1}\cdots Q_{n},\qquad p<n,
\end{equation}
$Q_{n,n}=\mathit{Id}$ and by convention $Q_{n+1,n}=\mathit{Id}$.

We now introduce our first two assumptions, which will be called upon
in the following.

\renewcommand\thelonglist{(H1)}
\renewcommand\labellonglist{\thelonglist}
\begin{longlist}
\item\label{hypdrift}
There exists $V\dvtx \mathsf{X}\rightarrow
[1,\infty)$
unbounded and constants $\delta>0$ and $\underline{d}\geq1$ with
the following properties. For each $d\in[\underline{d},\infty
)$
there exists $b_{d}<\infty$ such that the following multiplicative
drift condition holds:
%
%e17 #&#
\begin{equation}\label{eqAdrift}
\sup_{n\geq1}Q_{n} \bigl(e^{V} \bigr)
\leq e^{V (1-\delta
)+b_{d}\mathbb{I}_{C_{d}}},
\end{equation}
where $C_{d}:= \{ x\in\mathsf{X};V(x)\leq d \} $.
\end{longlist}
Whenever \ref{hypdrift} holds we may also consider:

\renewcommand\thelonglist{(H2)}
\renewcommand\labellonglist{\thelonglist}
\begin{longlist}
\item\label{hypmuv}
$\mu\in\mathcal{P}_{v}$, where
$v=e^{V}$ is as in \ref{hypdrift}.
\end{longlist}

We may now proceed with some further definitions. Define the collection
of measures $ (\gamma_{n};n\geq0 )$ and probability measures
$ (\eta_{n};n\geq0 )$
%
%e18 #&#
\begin{equation}\label{eqetafromQ}
\gamma_{n} (A ):=\mu Q_{0,n}(A),\qquad \eta_{n}(A):=
\frac
{\gamma_{n} (A )}{\gamma_{n} (1 )},\qquad A\in\mathcal{B}(\mathsf{X}),
\end{equation}
where the dependence of $ (\gamma_{n} )$ and $ (\eta
_{n} )$
on the initial distribution $\mu$ is suppressed from the notation.

Before going further we note the following elementary implications
of assumptions \ref{hypdrift} and \ref{hypmuv} introduced
so far. Assumption \ref{hypdrift} implies that for all $n\geq1$
and $x\in\mathsf{X}$, $Q_{n} (e^{V} )(x)/e^{V(x)}\leq
e^{b_{\underline{d}}}<\infty$
and thus for all $0\leq p\leq n$ and $x\in\mathsf{X}$,
%
%e19 #&#
\begin{equation}\label{eqQpn<infty}
Q_{p,n} \bigl(e^{V} \bigr) (x) < \infty.
\end{equation}
Combined with assumption \ref{hypmuv}, we also observe that for
all $n\geq0$, $\eta_{n}\in\mathcal{P}_{v}$.

It is straightforward to verify that the unnormalized measures $
(\gamma_{n} )$
have the following product representation:
%
%e20 #&#
\begin{equation}\label{eqgammaproduct}
\gamma_{n} (A )=\prod_{p=0}^{n-1}
\eta_{p} (G_{p} )\eta_{n} (A ),\qquad n\geq1.
\end{equation}

We denote by $\mathbb{E_{\mu}}$ the expectation w.r.t. to the canonical
law of the nonhomogeneous Markov chain $ (X_{n};n\geq0 )$
where $X_{0}\sim\mu$ and $X_{n}| \{ X_{n-1}=x_{n-1} \} \sim
M_{n} (x_{n-1},\cdot)$.
For $p\leq n$ and a suitable test function $\varphi$ we abuse notation
by writing
\[
\mathbb{E}_{p,x} \bigl[\varphi(X_{p},\ldots,X_{n} ) \bigr] := \mathbb{E}_{\mu} \bigl[\varphi
(X_{p},\ldots,X_{n} )|X_{p}=x \bigr],
\]
and for a probability measure $\eta$ we write
\[
\mathbb{E}_{p,\eta} \bigl[\varphi(X_{p},\ldots,X_{n} ) \bigr] := \int_{\mathsf{X}}\eta(dx )
\mathbb{E}_{p,x} \bigl[\varphi(X_{p},\ldots,X_{n}
) \bigr].
\]
Under these notational conventions we have, for $0\leq p<n$ and
$\eta\in\mathcal{P}$, the identity
\[
\eta Q_{p,n}(A)=\mathbb{E}_{p,\eta} \Biggl[\prod
_{q=p}^{n-1}G_{q} (X_{q} )
\mathbb{I} [X_{n}\in A ] \Biggr].
\]
In particular,
\[
\eta_{p}Q_{p,n}(A)=\mathbb{E}_{p,\eta_{p}} \Biggl[\prod
_{q=p}^{n-1}G_{q}
(X_{q} )\mathbb{I} [X_{n}\in A ] \Biggr]=\prod
_{q=p}^{n-1}\eta_{q} (G_{q} )
\eta_{n} (A )
\]
due to (\ref{eqetafromQ}) and (\ref{eqgammaproduct}), which
will be used repeatedly.
%
%de1 #&#
\begin{defn}[($\lambda$-values and $h$-functions)]
\label{defhfunctions}
For $n\geq0$ let
\[
\lambda_{n}:=\eta_{n} (G_{n} ),
\]
and for $0\leq p\leq n$ let $h_{n,p}\dvtx \mathsf{X}\rightarrow(0,\infty
)$
be the function defined by
%
%e21 #&#
\begin{equation}\label{eqhfuncdefn}
h_{n,n}(x):=1,\qquad h_{p,n}(x):=\frac{Q_{p,n} (1
)(x)}{\prod_{q=p}^{n-1}\lambda_{q}},\qquad p<n.
\end{equation}
\end{defn}
%
%re1 #&#
\begin{rem}
It is stressed that each $\lambda_{p}$, and therefore each $h_{p,n}$,
depends implicitly on the initial distribution $\mu$. With the exception
of Corollary \ref{Corfilterstability}, throughout the following
$\mu$ should be understood as arbitrary but fixed.

The two other main assumptions are the following.
\end{rem}
\renewcommand\thelonglist{(H3)}
\renewcommand\labellonglist{\thelonglist}
\begin{longlist}
\item\label{hypminoronly} With $\underline{d}$ as
in \ref{hypdrift}, for each $d\in[\underline{d},\infty)$,
\[
Q_{n} (x,C_{d} )>0 \qquad\forall x\in\mathsf{X}, n\geq1,
\]
and there exists $\varepsilon_{d}^{-}>0$ and $\nu_{d}\in\mathcal{P}_{v}$,
such that
\[
\inf_{n\geq1}Q_{n} (x,C_{d}\cap A ) \geq
\varepsilon_{d}^{-}\nu_{d} (C_{d}\cap A )\qquad
\forall x\in C_{d}, A\in\mathcal{B} (\mathsf{X} )
\]
with $\nu_{d} (C_{r} )>0$, for all $r\in[\underline
{d},d ]$.
\end{longlist}
When \ref{hypdrift} and \ref{hypminoronly} hold,
we may also consider:

\renewcommand\thelonglist{(H4)}
\renewcommand\labellonglist{\thelonglist}
\begin{longlist}
\item\label{hypminorandmaj} With $\underline{d}$
as in \ref{hypdrift} and $ (\nu_{d} )$, $ (\varepsilon
_{d}^{-} )$
as in \ref{hypminoronly}, for each $d\in[\underline{d},\infty
)$,
there exists $\varepsilon_{d}^{+}\in[\varepsilon_{d}^{-},\infty)$
such that
\[
\sup_{n\geq1}Q_{n} (x,C_{d}\cap A ) \leq
\varepsilon_{d}^{+}\nu_{d} (C_{d}\cap A )\qquad
\forall x\in C_{d}, A\in\mathcal{B} (\mathsf{X} ).
\]
\end{longlist}

%s4.1.1 #&#
\subsubsection{Comments on the assumptions}\label{subComments-on-the}

Assumptions \ref{hypminoronly}--\ref{hypminorandmaj} taken
together are more specific than the local-Doeblin condition of \citet
{filtertheDFMP09}
(when the latter is considered as holding for nonnegative kernels)
because they are phrased in terms of the level sets for $V$ and hold
time-simultaneously. It is possible to obtain results which are the
analogue of those presented herein under multi-step versions of
\ref{hypminoronly}--\ref{hypminorandmaj},
but this involves substantial notational complications which would
obscure presentation.

Assumption \ref{hypdrift} is a type of multiplicative drift condition
involving the Markov kernels $ (M_{n} )$ and the potential
functions $ (G_{n} )$. A notable characteristic of this
assumption is that it implies that for all $\varepsilon>0$ there exists
$d\geq\underline{d}$ such that $\interleave Q_{n}-\mathbb
{I}_{C_{d}}Q_{n}\interleave_{v}<\varepsilon$
for all $n\geq1$, which is itself a time-simultaneous version of
\citet{filtertheDFMP09}, condition H2.

In the above definitions the functions $ (G_{n} )$ have
been taken as strictly positive. It would be interesting to also consider
vanishing potential functions, but that situation is more complicated
as the particle system may become extinct.

%s4.1.2 #&#
\subsubsection{Particle system}

The particle system may be considered a canonical nonhomogeneous
Markov chain and therefore its definition is only sketched. For $N\geq1$,
and each $n\geq0$ let $\xi_{n}= (\xi_{n}^{1},\ldots,\xi
_{n}^{N} )$,
be a $\mathsf{X}^{N}$-valued and then define
\begin{eqnarray*}
\eta_{n}^{N}&:=&\frac{1}{N}\sum
_{i=1}^{N}\delta_{\xi_{n}^{i}},\qquad n\geq0,
\\
\gamma_{0}^{N}&:=&\eta_{0}^{N},
\\
\gamma_{n}^{N}&:=& \Biggl[\prod
_{p=0}^{n-1}\eta_{p}^{N}
(G_{p} ) \Biggr]\eta_{n}^{N},\qquad n\geq1.
\end{eqnarray*}
The particle system of population size $N$ is the
$\mathsf{X}^{N}$-valued Markov chain with transitions given
symbolically by
\[
\bigl(\xi_{0}^{1},\ldots,\xi_{0}^{N}
\bigr)\iid\mu,\qquad \bigl(\xi_{n}^{1},\ldots,
\xi_{n}^{N} \bigr)\big|\xi_{n-1} \iid
\frac{\eta
_{n-1}^{N}Q_{n} (\cdot)}{\eta_{n-1}^{N} (G_{n-1}
)},\qquad n\geq1.
\]

%re2 #&#
\begin{rem}\label{re2}
In order to obtain Algorithm \ref{algpf} take $G_{n}(x):=g(x,y_{n})$,
$M_{n}(x,\allowbreak dx'):=f(x,dx')$. In this case $\eta_{n}^{N}\equiv\pi_{n}^{N}$
and $\gamma_{n}^{N}(1)\equiv Z_{n}^N$, and similarly, $\eta_{n}\equiv\pi_{n}$,
$\gamma_{n}(1)\equiv Z_{n}$. Other particle filters [such as the
``fully-adapted'' auxiliary particle filter of \citet{pitt1999filtering}]
arise from other choices of $G_{n}$ and $M_{n}$. More generally,
the state-space $\mathsf{X}$ may be augmented, for example, to $\mathsf{X}^{2}$,
in order to accommodate $M_{n}$ corresponding to other choices of
proposal kernel and corresponding importance weight; see, for example,
\citet{doucet2000sequential}. In such cases one would need multi-step
versions of \ref{hypminoronly}--\ref{hypminorandmaj}.
\end{rem}

%s4.2 #&#
\subsection{Uniform $v$-controls}\label{subUniform--controls}

The main results of this section are Propositions \ref{propetaboundedinv}
and \ref{propequiv}, which establish uniform controls on the measures
$ (\eta_{n} )$, the $\lambda$-values and the $h$-functions.
The uniform bounds of these propositions feature centrally in the
proofs of the stability
results which then follow.

The first ingredient is the following lemma, which establishes
some relationships between the measures $ (\eta_{n} )$,
the $\lambda$-values and the $h$-functions.

%le1 #&#
\begin{lem}
\label{lemeigfuncmeas}Assume
\ref{hypdrift}--\ref{hypmuv}.
The measures $ (\eta_{n} )$, $h$-functions and $\lambda$-values
satisfy, for any $n\geq1$ and $0\leq p<n$, the recursive formulas
%
%e22 #&#
\begin{equation}\label{eqeigfuncmeas}
\eta_{p}Q_{p+1}=\lambda_{p}
\eta_{p+1},\qquad Q_{p+1} (h_{p+1,n} )=\lambda_{p}h_{p,n}
\end{equation}
and
\[
\eta_{p} (h_{p,n} )=1.
\]
Furthermore $h_{p,n}\in\mathcal{L}_{v}$ where $v=e^{V}$ is as in
\ref{hypdrift}.
\end{lem}
\begin{pf}
For the measure equation,
\[
\eta_{n} (A )=\frac{\gamma_{n} (A )}{\gamma_{n}
(1 )}=\frac{\gamma_{n-1}Q_{n} (A )}{\gamma_{n-1}Q_{n}
(1 )}=
\frac{\eta_{n-1}Q_{n} (A )}{\eta_{n-1}Q_{n}
(1 )}=\frac{\eta_{n-1}Q_{n} (A )}{\eta_{n-1}
(G_{n-1} )},
\]
where the third equality is due to the product formula (\ref{eqgammaproduct}).
For the $h$-function equation, using Definition \ref{defhfunctions},
\[
h_{p-1,n}=\frac{Q_{p-1,n} (1 )}{\prod_{q=p-1}^{n-1}\lambda
_{q}}=\frac{1}{\lambda_{p-1}}\frac{Q_{p}Q_{p,n} (1 )}{\prod
_{q=p}^{n-1}\lambda_{q}}=
\frac{1}{\lambda_{p-1}}Q_{p} (h_{p,n} ).
\]
The equality $\eta_{p} (h_{p,n} )=1$ is direct from (\ref
{eqetafromQ})
and the definition of $h_{p,n}$. The assertion $h_{p,n}\in\mathcal{L}_{v}$
follows immediately from Definition \ref{defhfunctions} and (\ref
{eqQpn<infty}).
\end{pf}
The second ingredient is the collection of kernels and drift functions
identified in the following definition (that these kernels are Markov
is a consequence of Lemma \ref{lemeigfuncmeas}).
%
%de2 #&#
\begin{defn}[($S$-kernels and drift functions)]
\label{deftwisted} For
$n\geq1$, $1\leq p\leq n$ let $S_{p,n}\dvtx \mathsf{X}\times\mathcal{B}
(\mathsf{X} )\rightarrow\mathbb{R}_{+}$
be the Markov kernel defined by
%
%e23 #&#
\begin{equation}\label{eqSpndefn}
S_{p,n} (x,A ):=\frac{Q_{p} (\mathbb{I}_{A}h_{p,n}
)(x)}{\lambda_{p-1}h_{p-1,n}(x)},
\end{equation}
and let $v_{p,n}\dvtx \mathsf{X}\rightarrow[1.\infty)$ be defined by
\[
v_{p,n} (x ):=\frac{v(x)}{h_{p,n}(x)}\llVert h_{p,n}\rrVert
_{v},
\]
where $v$ is as in \ref{hypdrift}.

For each $n\geq1$ and $\eta\in\mathcal{P}$, we denote by $\check
{\mathbb{E}}_{\eta}^{(n)}$
expectation w.r.t. the canonical law of the $ (n+1 )$-step
nonhomogeneous Markov chain $ \{ \check{X}_{p,n};0\leq p\leq
n \} $
with $\check{X}_{0,n}\sim\eta$ and for $1\leq p\leq n$,
$\check{X}_{p,n}| \{ \check{X}_{p-1,n}=\check{x}_{p-1,n} \}
\sim S_{p,n} (\check{x}_{p-1,n},\cdot)$.
By analogy to the definitions of Section \ref{subFKsetup}, for
each $n\geq1$ we write
\[
\check{\mathbb{E}}_{p,x}^{(n)} \bigl[\varphi(
\check{X}_{p,n},\ldots,\check{X}_{n,n} ) \bigr] :=
\check{\mathbb{E}}_{\eta}^{(n)} \bigl[\varphi(
\check{X}_{p,n},\ldots,\check{X}_{n,n} )|\check
{X}_{p,n}=x \bigr].
\]

The $S$-kernels and the corresponding expectations are of interest
due to the following change-of-measure identity.
\end{defn}
%
%le2 #&#
\begin{lem}
\label{lemchangeofmeasure} Assume \ref{hypdrift}--\ref{hypmuv}.
For any $n\geq1$, $0\leq p<n$, a suitable test function $\varphi$
and $x\in\mathsf{X}$,
\[
\frac{\mathbb{E}_{p,x} [\prod_{q=p}^{n-1}G_{q} (X_{q}
)\varphi(X_{p},\ldots,X_{n} ) ]}{\mathbb{E}_{p,\eta
_{p}} [\prod_{q=p}^{n-1}G_{q} (X_{q} ) ]} = h_{p,n}(x)\check
{\mathbb{E}}_{p,x}^{(n)}
\bigl[\varphi(\check{X}_{p,n},\ldots,\check{X}_{n,n} )
\bigr].
\]
\end{lem}
\begin{pf}
From Definitions \ref{defhfunctions} and \ref{deftwisted},
\begin{eqnarray*}
& & \frac{\mathbb{E}_{p,x} [\prod_{q=p}^{n-1}G_{q} (X_{q}
)\varphi(X_{p},\ldots,X_{n} ) ]}{\eta_{p}Q_{p,n}
(1 )}
\\
& &\qquad =h_{p,n}(x)\mathbb{E}_{p,x} \Biggl[\prod
_{q=p}^{n-1}\frac{G_{q}
(X_{q} )}{\lambda_{q}}\frac{h_{q+1,n} (X_{q+1}
)}{h_{q,n} (X_{q} )}
\varphi(X_{0},\ldots,X_{n} )\frac
{1}{h_{n,n} (X_{n} )} \Biggr]
\\
& &\qquad =h_{p,n}(x)\check{\mathbb{E}}_{p,x}^{(n)}
\bigl[\varphi(\check{X}_{p,n},\ldots,\check{X}_{n,n} )
\bigr].
\end{eqnarray*}
\upqed\end{pf}
%
%re3 #&#
\begin{rem}
The $S$-kernels have previously been identified as playing an important
role when analyzing stability properties of Feynman--Kac formulas and
particle systems; see \citet{smctheDMG01}, albeit written in a
slightly different form. From Definition \ref{defhfunctions} we
have immediately that
\[
S_{p,n} (x,A )=\frac{Q_{p} (\mathbb{I}_{A}h_{p,n}
)(x)}{\lambda_{p-1}h_{p-1,n}(x)}=\frac{Q_{p} (\mathbb
{I}_{A}Q_{p,n}(1) )(x)}{Q_{p-1,n}(1)(x)},
\]
and it is in the latter form that these kernels are usually considered.
However, in the context of the Lyapunov drift techniques employed
here, (\ref{eqSpndefn}) expressed in terms of the $\lambda$-values
and $h$-functions plays a central role in proofs of the two following
propositions. The main theme of the proof of Proposition \ref
{propetaboundedinv}
is to obtain uniform bounds on $\llVert\eta_{n}\rrVert_{v}$
via the representation of Lemma \ref{lemchangeofmeasure}, the
identity $\check{\mathbb{E}}_{p,x}^{(n)} [v (\check
{X}_{n,n} ) ]=S_{p+1,n}\cdots S_{n,n} (v )(x)$
and the drift functions $ (v_{p,n} )$.

Note that Proposition \ref{propetaboundedinv} does not require
the majorization-type assumption \ref{hypminorandmaj}.
\end{rem}
%
%pr1 #&#
\begin{prop}
\label{propetaboundedinv}Assume \ref{hypdrift}--\ref{hypminoronly},
and let $v$ be as therein. Then there exists a finite constant $c_{\mu}$
depending on $\mu$ and the quantities in \ref{hypdrift} and
\ref{hypminoronly}, such that
\[
\sup_{n\geq0}\llVert\eta_{n}\rrVert_{v}
\leq c_{\mu}\mu(v ).
\]
\end{prop}
\begin{pf}
See Section \ref{secProofs}.
\end{pf}
The interest in the uniform bound of Proposition \ref{propetaboundedinv}
is that, via the following proposition, we obtain some uniform bounds
on the $\lambda$-values and $h$-functions.
%
%pr2 #&#
\begin{prop}
\label{propequiv} Assume \ref{hypdrift}--\ref{hypminoronly},
and let $v$ be as therein. Then (1)--(2) below are equivalent:

(1) $\sup_{n\geq0}\llVert\eta_{n}\rrVert
_{v}<\infty$;

(2) $\inf_{n\geq0}\lambda_{n}>0$;

\textup{If additionally \ref{hypminorandmaj} holds, then} (1)
and (2) are equivalent to (3),

{}(3) $\sup_{n\geq1}\sup_{0\leq p\leq n}\llVert h_{p,n}\rrVert
_{v}<\infty$.
\end{prop}
\begin{pf}
Lemmata \ref{lemtightness12}, \ref{lemtightness2to3} and
\ref{lemhpnboundedinv}. See Section \ref{secProofs}.
\end{pf}
Before proceeding further, note that in the results from this point
on, the statements often feature a constant $c_{\mu}$. The value
of this constant may change from one result to the next.

%s4.3 #&#
\subsection{A multiplicative stability theorem}\label{subMET}

The form of the following result can be interpreted as a nonhomogeneous
analogue of the multiplicative ergodic theorem of \citet{mctheKM05}
in the context of positive operators, for direct comparison the reader
is referred to \citet{whiteley2011}, Theorem~2.2, equation (2.9).
This proposition will be applied in Section \ref{subVariance-bounds}
to bound the asymptotic variance associated with $ (\eta
_{n}^{N} )$.
The proof is postponed.
%
%th1 #&#
\begin{theorem}
\label{thmMET}Assume \ref{hypdrift}--\ref{hypminorandmaj}.
Then there exists $\rho<1$ depending on $\mu$ and the constants
in \ref{hypdrift}, \ref{hypminoronly} and \ref{hypminorandmaj}
and $c_{\mu}<\infty$ depending on $\mu$ and the quantities in
\ref{hypdrift}--\ref{hypminorandmaj}
such that for any $\varphi\in\mathcal{L}_{v}$, $n\geq1$ and $0\leq p<n$,
\[
\biggl\llvert\frac{Q_{p,n} (\varphi)(x)}{\prod_{q=p}^{n-1}\lambda
_{q}}-h_{p,n}(x)\eta_{n} (
\varphi)\biggr\rrvert\leq\rho^{n-p}\llVert\varphi\rrVert
_{v}c_{\mu}v(x)\mu(v )\qquad \forall x\in\mathsf{X}.
\]
\end{theorem}
\begin{pf}
See Section \ref{secProofs}.
\end{pf}
As a consequence of this theorem we obtain $v$-norm exponential stability
with respect to initial condition for measures $ (\eta_{n} )$.
%
%co1 #&#
\begin{cor}
\label{Corfilterstability}Assume \ref{hypdrift}--\ref
{hypminorandmaj}, and
then with $\rho$ and $\mu$ as in Theorem \ref{thmMET}. For any
$\mu^{\prime}\in\mathcal{P}_{v}$, there exists $c_{\mu,\mu'}<\infty$
such that
\[
\bigl\llVert\eta_{n}^{(\mu)}-\eta_{n}^{(\mu^{\prime})}
\bigr\rrVert_{v} \leq \rho^{n}c_{\mu,\mu^{\prime}}\mu(v )
\mu^{\prime} (v ),
\]
where $\eta_{n}^{ (\mu)}:=\frac{\mu Q_{0,n}}{\mu Q_{0,n}
(1 )}$
and $\eta_{n}^{ (\mu^{\prime} )}:=\frac{\mu^{\prime
}Q_{0,n}}{\mu^{\prime}Q_{0,n} (1 )}$.
\end{cor}
\begin{pf}
Taking the bound of Theorem \ref{thmMET} and integrating w.r.t.
$\mu'$ gives
\[
\biggl\llvert\frac{\mu^{\prime}Q_{0,n} (\varphi)}{\prod
_{p=0}^{n-1}\lambda_{p}}-\mu^{\prime} (h_{0,n} )
\eta_{n}^{
(\mu)} (\varphi)\biggr\rrvert \leq
\rho^{n}\llVert\varphi\rrVert_{v}c_{\mu}\mu(v
)\mu^{\prime} (v ).
\]
It is stressed that in the above display $\lambda_{p}$ and $h_{0,n}$
are as in Definition \ref{defhfunctions}, that is, dependent on $\mu$,
but not on $\mu^{\prime}$. Now as $\mu^{\prime}\in\mathcal{P}_{v}$,
for any $d\in[\underline{d},\infty)$, $\mu^{\prime}
(C_{d}^{c} )\leq\mu^{\prime} (\mathbb{I}_{C_{d}^{c}}e^{V}
)/e^{d}\leq\mu(e^{V} )/e^{d}$
so there exists $d\in[\underline{d},\infty)$ such that\vadjust{\goodbreak}
$\mu^{\prime} (C_{d} )>0$. Then dividing through by $\mu
^{\prime} (h_{0,n} )=\mu^{\prime}Q_{0,n}(1)/\prod_{p=0}^{n-1}\lambda_{p}$,
\begin{eqnarray*}
\biggl\llvert\frac{\mu^{\prime}Q_{0,n} (\varphi)}{\mu^{\prime
}Q_{0,n} (1 )}-\eta_{n}^{ (\mu)} (\varphi)
\biggr\rrvert& \leq& \rho^{n}\llVert\varphi\rrVert_{v}
\frac{c_{\mu
}}{\mu^{\prime} (h_{0,n} )}\mu(v )\mu^{\prime} (v )
\\
& \leq& \rho^{n}\llVert\varphi\rrVert_{v}
\frac{c_{\mu}}{\mu
^{\prime} (C_{d} )\inf_{x\in C_{d}}h_{0,n}(x)}\mu(v )\mu^{\prime} (v )
\\
& \leq& \rho^{n}\llVert\varphi\rrVert_{v}c_{\mu,\mu^{\prime
}}
\mu(v )\mu^{\prime} (v ),
\end{eqnarray*}
where the final inequality holds due to Lemma \ref
{lemhpnboundedbelowonC}.\vspace*{-2pt}
\end{pf}

%s4.4 #&#
\subsection{Exponential moments for additive functionals}
\label{subExponential-moments-for}

We now present a result on finite exponential moments for a class
of additive, possibly unbounded path space functionals. It will be
applied in Section \ref{subVariance-bounds} to bounds on the relative
variance associated with $\gamma_{n}^{N}(1)$. The proof is mostly
technical and is given in Section \ref{secProofs}.\vspace*{-2pt}
%
%th2 #&#
\begin{theorem}
\label{thmexpmoments} Assume \ref{hypdrift}--\ref{hypminorandmaj},
and let $\delta$ and $v$ be as therein. Then there exists a finite
constant $c_{\mu}$ depending on $\mu$ and the quantities in
\ref{hypdrift}--\ref{hypminorandmaj}
such that for any collection of measurable functions $ \{
F_{n};n\geq1 \} $
with each $F_{n}\dvtx \mathsf{X}\rightarrow\mathbb{R}$ and $\sup_{x}
(\llvert F_{n}(x)\rrvert-\delta V(x) )<\infty$;
any $n\geq1$, $0\leq s\leq n+1$, and $ (i_{1},\ldots,i_{s} )\in
\mathcal{I}_{n,s}$,
\begin{eqnarray*}
&&\frac{\mathbb{E}_{\mu} [\prod_{p=0}^{n-1}G_{p} (X_{p} )\exp
(\sum_{k\in\{ i_{1},\ldots,i_{s} \} }\llvert F_{k}
(X_{k} )\rrvert ) ]}{\mathbb{E}_{\mu} [\prod_{p=0}^{n-1}G_{p} (X_{p} ) ]}
\\
&&\qquad\leq c_{\mu}^{s}\mu(v )\prod_{k\in\{ i_{1},\ldots,i_{s} \} }
\bigl\llVert e^{\llvert F_{k}\rrvert}\bigr\rrVert_{v^{\delta}}
\end{eqnarray*}
with the conventions that, when $s=0$, the summation on the left-hand
side is zero, and the product on the right-hand side is unity.\vspace*{-2pt}
\end{theorem}
\begin{pf}
See Section \ref{secProofs}.\vspace*{-2pt}
\end{pf}

%s4.5 #&#
\subsection{Variance bounds}\label{subVariance-bounds}\vspace*{-2pt}
%
%re4 #&#
\begin{rem}
At this point we introduce a further assumption, \ref{hypGbounded}
below. This assumption is not necessary for all of the results of
this section, but is employed for the following three reasons: (1) it
is not too restrictive in filtering applications; (2) it allows Lemma
\ref{lemtransferdrift}
below to be invoked [an equivalent result can also be obtained without
\ref{hypGbounded}, but subject to constraints on the growth
rates of $ (G_{n} )$ and the assumption that the Markov
kernels $ (M_{n} )$ themselves obey a suitable simultaneous
multiplicative drift condition]; (3) it allows an existing CLT
for particle systems to be simply stated below without proof; see
also Remark \ref{remark5}.
\end{rem}

\renewcommand\thelonglist{(H5)}
\renewcommand\labellonglist{\thelonglist}
\begin{longlist}
\item\label{hypGbounded} $\sup_{n\geq0}\sup_{x\in\mathsf
{X}}G_{n}(x)<\infty$.
\end{longlist}\eject

The following lemma plays an important technical role in the variance
results which follow.
%
%le3 #&#
\begin{lem}
\label{lemtransferdrift}Assume \ref{hypdrift}--\ref{hypGbounded}
with $v$ the drift function in \ref{hypdrift}--\ref{hypminorandmaj}.
Then for any $\alpha\in(0,1)$, the statements of
\ref{hypdrift}--\ref{hypminorandmaj}
also hold for the drift function $v_{1}:=v^{\alpha}$ and with $\alpha$-dependent
constants.
\end{lem}
\begin{pf}
Let $\bar{G}:=\sup_{n\geq0}\sup_{x\in\mathsf{X}}G_{n}(x)$. Then for
all $x\in\mathsf{X}$ and any $d\in[\underline{d},\infty)$
as in \ref{hypdrift},
\begin{eqnarray*}
\sup_{n\geq1}Q_{n} \bigl(e^{\alpha V} \bigr) (x)
& \leq& \bar{G}\sup_{n\geq1} \biggl[\frac{G_{n-1}(x)}{\bar{G}}M_{n}
\bigl(e^{V} \bigr) (x) \biggr]^{\alpha}
\\
& = & \bar{G}^{1-\alpha}\sup_{n\geq1} \bigl[Q_{n}
\bigl(e^{V} \bigr) (x) \bigr]^{\alpha}
\\
& \leq& \exp\bigl[\alpha V(x) (1-\delta)+\alpha b_{d}\mathbb
{I}_{C_{d}}(x)+(1-\alpha)\log\bar{G} \bigr],
\end{eqnarray*}
where Jensen's inequality and \ref{hypdrift} have been applied,
and $\delta$, $b_{d}$ and $C_{d}= \{ x\in\mathsf{X};V(x)\leq d
\} $
are as in \ref{hypdrift}. Then for any $\delta_{0}\in(0,\delta
)$
and $\bar{G}<\infty$ there exists $\underline{d}_{\alpha}\in
[\underline{d},\infty)$
such that for any $d\in[\underline{d}_{\alpha},\infty)$
and $x\notin\{ x\in\mathsf{X};\alpha V(x)\leq d \} $,
%
%e24 #&#
\begin{eqnarray}\label{eqQalphadrift1}\quad
\sup_{n\geq1}Q_{n} \bigl(e^{\alpha V} \bigr) (x)
& \leq& \exp\bigl[\alpha V(x) (1-\delta_{0} )-\alpha d (\delta-
\delta_{0} )+(1-\alpha)\log\bar{G} \bigr]
\nonumber\\[-8pt]\\[-8pt]
& \leq& \exp\bigl[\alpha V(x) (1-\delta_{0} ) \bigr]\nonumber
\end{eqnarray}
and for $x\in\{ x\in\mathsf{X};\alpha V(x)\leq d \} $,
%
%e25 #&#
\begin{eqnarray}\label{eqQalphadrift2}
\sup_{n\geq1}Q_{n} \bigl(e^{\alpha V} \bigr) (x)
& \leq& \exp\bigl[\alpha d(1-\delta)+\alpha b_{d}+(1-\alpha)\log
\bar{G} \bigr]
\nonumber\\[-8pt]\\[-8pt]
& =:&  \exp(b_{d,\alpha} ).\nonumber
\end{eqnarray}
The statement of \ref{hypdrift} holds with the drift function
$v_{1}:=v^{\alpha}$ because equations (\ref{eqQalphadrift1})--(\ref
{eqQalphadrift2})
show that we may replace $\underline{d}$, $\delta$, $b_{d}$, $C_{d}$
in the corresponding statements with $\underline{d}_{\alpha}$, $\delta_{0}$,
$b_{d,\alpha}$, $\{ x\in\mathsf{X};\alpha V(x)\leq d \} $,
respectively.

It is immediate that \ref{hypmuv} holds for $v^{\alpha}$
because $v\geq1$. \ref{hypminoronly}--\ref{hypminorandmaj}
also hold for~$v^{\alpha}$, by replacing $\underline{d}$,
$C_{d}$, $\varepsilon_{d}^{-}$,
$\varepsilon_{d}^{+}$, $\nu_{d}$ with $\underline{d}_{\alpha}$, $ \{
x\in\mathsf{X};\alpha V(x)\leq d \} $,
$\varepsilon_{d/\alpha}^{-}$, $\varepsilon_{d/\alpha}^{+}$, $\nu_{d/\alpha}$,
respectively.
\end{pf}

%s4.5.1 #&#
\subsubsection{\texorpdfstring{Asymptotic variance for $\eta_{n}^{N}$}
{Asymptotic variance for eta n N}}\label{subAsymptotic-variance-for}
%
%re5 #&#
\begin{rem}\label{remark5}
There are several existing CLT results for the particle systems in
question; see, for example, \citet{smctheC04,smctheDM08}. We choose
to present that of Del Moral [(\citeyear{smctheoryDm04}), Proposition 9.4.2], as
it holds immediately under \ref{hypGbounded}, and we may state
also the corresponding asymptotic variance expression with essentially
no further work. The restriction is that the stated result holds only
for bounded functions. It is of interest\vadjust{\goodbreak} to investigate whether the
same result holds for a suitable class of possibly unbounded functions
in terms of $v$, for example, via the techniques of \citet{smctheC04}
or \citet{smctheDM08}, but this is beyond the scope of the present
article.

The following CLT holds for errors associated with the particle approximation
measures $ (\eta_{n}^{N} )$. Straightforward manipulations
of the asymptotic variance expression of Del Moral [(\citeyear{smctheoryDm04}),
Proposition 9.4.2]
show that it can be written as in (\ref{eqasympvar}) below, in
terms of the $h$-functions and $\lambda$-values.
\end{rem}
%
%th3 #&#
\begin{theorem}[{[\citet{smctheoryDm04}, Proposition 9.4.2]}]
\label{thmCLT} Assume
\ref{hypGbounded}. Then for $\varphi\dvtx \mathsf{X}\rightarrow\mathbb{R}$
bounded and measurable and any $n\geq1$,
\[
\sqrt{N} \bigl(\eta_{n}^{N}-\eta_{n} \bigr) (
\varphi) \rightarrow \mathcal{N} \bigl(0,\sigma_{n}^{2}
\bigr)
\]
in distribution as $N\rightarrow\infty$, where
%
%e26 #&#
\begin{equation}\label{eqasympvar}
\sigma_{n}^{2}:=\eta_{n} \bigl[ \bigl(\varphi-
\eta_{n} (\varphi) \bigr)^{2} \bigr]+\sum
_{p=0}^{n-1}\eta_{p} \biggl[ \biggl(
\frac
{Q_{p,n} (\varphi)}{\prod_{q=p}^{n-1}\lambda_{q}}-h_{p,n}\eta_{n}
(\varphi)
\biggr)^{2} \biggr].
\end{equation}
\end{theorem}

We can readily apply the result of Theorem \ref{thmMET} to obtain
a time-uniform bound on the asymptotic variance.
%
%pr3 #&#
\begin{prop}
\label{propvarianceunibound}Assume \ref{hypdrift}--\ref{hypGbounded}.
Then there exists $c_{\mu}<\infty$ depending only on $\mu$ and the
quantities in \ref{hypdrift}--\ref{hypGbounded} such that
for any $n\geq1$,
\[
\sigma_{n}^{2}\leq\eta_{n} \bigl[ \bigl(
\varphi-\eta_{n} (\varphi) \bigr)^{2} \bigr]+c_{\mu}
\llVert\varphi\rrVert_{1}^{2}\mu(v )^{2},
\]
where $v$ is as in \ref{hypdrift} and $\varphi$ and $\sigma_{n}^{2}$
are as in Theorem \ref{thmCLT}.
\end{prop}
\begin{pf}
As \ref{hypdrift}--\ref{hypminorandmaj} are assumed to
hold with some drift function $v$, then by Lemma \ref{lemtransferdrift},
the same assumptions hold with the drift function $v^{1/2}$ and suitable
constants. Then applying Theorem \ref{thmMET} [using the drift $v^{1/2}$
and the corresponding instances \ref{hypdrift}--\ref{hypminorandmaj}]
and then Proposition \ref{propetaboundedinv} (using the drift
$v$), we find that there is $c_{\mu}<\infty$ such that
\begin{eqnarray*}
\eta_{p} \biggl[ \biggl(\frac{Q_{p,n} (\varphi)}{\prod
_{q=p}^{n-1}\lambda_{q}}-h_{p,n}
\eta_{n} (\varphi) \biggr)^{2} \biggr] & \leq&
\rho^{2(n-p)}c_{\mu}\llVert\varphi\rrVert_{1}^{2}
\mu\bigl(v^{1/2} \bigr)^{2}\eta_{p} (v )
\\
& \leq& \rho^{2(n-p)}\llVert\varphi\rrVert_{1}^{2}c_{\mu}
\mu(v )^{2},
\end{eqnarray*}
and the statement of the theorem follows by summing.
\end{pf}

%s4.5.2 #&#
\subsubsection{\texorpdfstring{Nonasymptotic variance for $\gamma_{n}^{N}(1)$}{Nonasymptotic variance for gamma n N(1)}}

For $n\geq1$ and $1\leq s\leq n+1$, define
\[
\Upsilon_{n}^{(i_{1},\ldots,i_{s})}:=\frac{\mu Q_{0,i_{1}}(1)\mathbb{E}_{\mu
} [\prod_{p=0}^{n-1}G_{p} (X_{p} )\prod_{j=1}^{s}Q_{i_{j},i_{j+1}}(1)
(X_{i_{j}} ) ]}{
[\gamma_{n} (1 ) ]^{2}}
\]
with the convention that $i_{s+1}=n$.\eject

Building from \citet{smcthedMPR09,smctheCdMG11} obtained
a nonasymptotic functional expansion of the relative variance associated
with $\gamma_{n}^{N}(1)$. Elementary manipulations of this relative
variance show that it may be written in terms of the quantities $
(\Upsilon_{n}^{(i_{1},\ldots,i_{s})} )$
as follows, and as we assume \ref{hypGbounded}, the quantities
involved are well defined [although this is not a necessary condition,
one may alternatively assume \ref{hypdrift}--\ref{hypmuv}].
%
%th4 #&#
\begin{theorem}[{[\citet{smctheCdMG11}, Proposition 3.4]}]
\label{thmCerou} Assume \ref{hypGbounded}. Then for any $n\geq1$,
\begin{eqnarray*}
& & \mathbb{E}_{\mu} \biggl[ \biggl(\frac{\gamma_{n}^{N} (1
)}{\gamma_{n}(1)}-1
\biggr)^{2} \biggr]
\\
& &\qquad =\sum_{s=1}^{n+1} \biggl(1-
\frac{1}{N} \biggr)^{ (n+1
)-s}\frac{1}{N^{s}}\sum
_{ (i_{1},\ldots,i_{s} )\in\mathcal
{I}_{n,s}} \bigl[\Upsilon_{n}^{(i_{1},\ldots,i_{s})}-1 \bigr],
\end{eqnarray*}
where the expectation is with respect to the law of the $N$-particle
system initialized from $\mu$.
\end{theorem}
We may now apply Theorem \ref{thmexpmoments} in order to obtain
the following linear-in-$n$ bound on the relative variance.
%
%pr4 #&#
\begin{prop}
\label{propvariance} Assume \ref{hypdrift}--\ref{hypGbounded}
and let $v$ be as therein. Then there exists a finite constant $c_{\mu}$
depending on $\mu$ and the quantities in \ref{hypdrift}--\ref{hypGbounded}
such that for any $n\geq1$,
\[
N>c_{\mu} (n+1 ) \quad\Longrightarrow\quad\mathbb{E}_{\mu} \biggl[
\biggl(\frac{\gamma_{n}^{N} (1 )}{\gamma
_{n}(1)}-1 \biggr)^{2} \biggr]\leq c_{\mu}
\frac{4}{N} (n+1 )\mu(v )^{2}.
\]
\end{prop}
\begin{pf}
Throughout the proof $c$ is a finite constant depending on $\mu$
and the quantities in \ref{hypdrift}--\ref{hypGbounded}
whose value may change on each appearance.

First notice that by Definition \ref{defhfunctions} and the product
formula (\ref{eqgammaproduct}) we may write
%
%e27 #&#
\begin{eqnarray}\label{equpsilonh}
\Upsilon_{n}^{(i_{1},\ldots,i_{s})} & = & \frac{\mu Q_{0,i_{1}}(1)}{\mu
Q_{0,i_{1}}(1)}
\frac{1}{\gamma_{n} (1 )}\mathbb{E}_{\mu} \Biggl[\prod
_{p=0}^{n-1}G_{p} (X_{p} )
\prod_{j=1}^{s} \biggl(\frac
{Q_{i_{j},i_{j+1}}(1) (X_{i_{j}} )}{\prod_{k=i_{j}}^{i_{j+1}-1}\lambda_{k}}
\biggr) \Biggr]
\nonumber\\[-8pt]\\[-8pt]
& = & \frac{1}{\gamma_{n} (1 )}\mathbb{E}_{\mu} \Biggl[\prod
_{p=0}^{n-1}G_{p} (X_{p} )
\prod_{j=1}^{s}h_{i_{j},i_{j+1}}
(X_{i_{j}} ) \Biggr]\nonumber
\end{eqnarray}
with the convention that $\prod_{n}^{n-1}=1$ in the first equality
to deal with the case $i_{s}=n$.

Let $v$ and $\delta$ be as in \ref{hypdrift}. Then by Lemma
\ref{lemtransferdrift}, the statements of \ref{hypdrift}--\ref
{hypminorandmaj}
also hold for the drift function $v^{\delta}$\vadjust{\goodbreak} and with constants
which depend on $\delta$. Then Propositions \ref{propetaboundedinv}
and \ref{propequiv} both applied with the drift function
$v^{\delta}$ and the corresponding instances of
\ref{hypdrift}--\ref{hypminorandmaj}
of show that
\[
\sup_{n\geq1}\sup_{0\leq p\leq n}\llVert
h_{p,n}\rrVert_{v^{\delta}}<\infty,
\]
so that, using representation (\ref{equpsilonh}), and applying
Theorem \ref{thmexpmoments} with the drift function $v$ and the
corresponding instances of \ref{hypdrift}--\ref{hypminorandmaj},
there exists a finite constant $c$ such that
\begin{eqnarray*}
\Upsilon_{n}^{(i_{1},\ldots,i_{s})} & \leq& c^{s}
\frac{1}{\gamma_{n}
(1 )}\mathbb{E}_{\mu} \Biggl[\prod
_{p=0}^{n-1}G_{p} (X_{p} )
\prod_{j=1}^{s}v^{\delta}
(X_{i_{j}} ) \Biggr]
\\
& \leq& c^{s}\mu(v).
\end{eqnarray*}
Therefore by Theorem \ref{thmCerou},
\[
\mathbb{E}_{\mu} \biggl[ \biggl(\frac{\gamma_{n}^{N} (1 )}{\gamma
_{n}(1)}-1
\biggr)^{2} \biggr]\leq\mu(v )^{2}\sum
_{s=1}^{n+1} \biggl(1-\frac{1}{N}
\biggr)^{ (n+1 )-s}\frac
{1}{N^{s}}\sum_{ (i_{1},\ldots,i_{s} )\in\mathcal{I}_{n,s}}c^{s}.
\]
The remainder of the proof then follows by the same arguments as
C{\'e}rou, Del Moral and Guyader
[(\citeyear{smctheCdMG11}), proofs of Theorem 5.1 and Corollary 5.2],
so the details are omitted.
\end{pf}

%s5 #&#
\section{Proofs and auxiliary results}\label{secProofs}

\subsection*{Auxiliary result for Section \protect\ref{subergsignalmodel}}
%
%le4 #&#
\begin{lem}
\label{lemARsignalmodel}When $f$ is the transition kernel corresponding
to the model of equations (\ref{eqKVmodel})--(\ref{eqKVmodelsigma}),
there exist $\underline{d}<\infty$ and $\delta>0$ such that, for
any $d\in[\underline{d},\infty)$, there exist $b_{d}<\infty$
and
%
%e28 #&#
\begin{equation}\label{eqfrontfdrift}
\int_{\mathsf{X}}f \bigl(x,dx^{\prime} \bigr)v
\bigl(x^{\prime}\bigr)\leq v(x)^{1-\delta}\exp\bigl[b_{d}
\mathbb{I}_{C_{d}}(x) \bigr],\qquad x\in\mathsf{X},
\end{equation}
where $v(x):=\exp(1+c\llvert x\rrvert )$ for $c$ a positive
constant, and furthermore for each such $d$ there exists $0<\varepsilon
_{d}^{-}<\varepsilon_{d}^{+}<\infty$
such that
%
%e29 #&#
\begin{equation}\label{eqKVmodlefmajmin}
\varepsilon_{d}^{-}\nu_{d} (A\cap C_{d} )
\leq f (x,A\cap C_{d} )\leq\varepsilon_{d}^{+}
\nu_{d} (A\cap C_{d} ),\qquad x\in C_{d},A\in\mathcal{B} (
\mathsf{X} ),\hspace*{-28pt}
\end{equation}
with $\nu_{d}$ the normalized restriction of Lebesgue measure to
$C_{d}$. Furthermore $\int_{C_{d}}f(x,dx')>0$, $\forall x\in\mathsf{X}$.
\end{lem}
\begin{pf}
As per \citet{filtertheDFMP09}, under the assumptions on the model,
there exists $\beta<\infty$ such that
\begin{eqnarray*}
\frac{\int_{\mathsf{X}}f (x,dx^{\prime} )v(x^{\prime})}{v(x)} & \leq&
\beta\exp\bigl[c \bigl(\bigl\llvert x+B(x)\bigr
\rrvert-\llvert x\rrvert\bigr) \bigr]
\\
& = & \beta\exp\biggl[-c\llvert x\rrvert\biggl(1-\frac{\llvert
x+B(x)\rrvert}{\llvert x\rrvert} \biggr)
\biggr],
\end{eqnarray*}
and then using (\ref{eqKVmodelb}), there exists $\delta_{1}>0$
such that for $\llvert x\rrvert$ sufficiently large,
\[
\biggl(1-\frac{\llvert x+B(x)\rrvert}{\llvert x\rrvert} \biggr)\geq
\delta_{1},
\]
so for such $\llvert x\rrvert$ and $\delta\in(0,\delta_{1})$,
\[
\frac{\int_{\mathsf{X}}f (x,dx^{\prime} )v(x^{\prime
})}{v(x)}\leq\exp\bigl[-V(x)\delta-c\llvert x\rrvert(\delta
_{1}-\delta)+\log\beta+1 \bigr],
\]
and by increasing $\llvert x\rrvert$ further if necessary, we conclude
that the result holds with $b_{d}:=d+\log\beta$. (\ref{eqKVmodlefmajmin})
and $\int_{C_{d}}f(x,dx')>0$ hold immediately.
\end{pf}

\subsection*{Proofs and results for Section \protect\ref{subUniform--controls}}

The proof of Proposition \ref{propetaboundedinv} is given after
Lemmas \ref{lemSdrift} and \ref{lemdriftbound}.
%
%le5 #&#
\begin{lem}
\label{lemSdrift} Assume \ref{hypdrift}--\ref{hypminoronly}.
Then for any $d\in[\underline{d},\infty)$, any $n\geq1$
and $1\leq p\leq n$, the following inequalities hold:
%
%e30 #&#
\begin{equation}\label{eqSpndrift}
S_{p,n} (v_{p,n} ) \leq \rho_{p,n}v_{p-1,n}+B_{p,n}
\mathbb{I}_{C_{d}},
\end{equation}
where
%
%e31 #&#
%e32 #&#
\begin{eqnarray}
\label{eqrhopndefn}
\rho_{p,n} &:= & \frac{e^{-\delta d}}{\lambda_{p-1}}\frac{\llVert
h_{p,n}\rrVert_{v}}{\llVert h_{p-1,n}\rrVert_{v}}<\infty,
\\
\label{eqBpndefn}
B_{p,n} &:= & \frac{e^{d(1-\delta)+b_{d}}}{\varepsilon_{d}^{-}}\llVert
h_{p,n}\rrVert
_{v}\frac{1}{\nu_{d} (\mathbb
{I}_{C_{d}}h_{p,n} )}<\infty
\end{eqnarray}
and with the dependence of $\rho_{p,n}$ and $B_{p,n}$ on $d$ suppressed
from the notation.
\end{lem}
\begin{pf}
For $x\notin C_{d}$,
\begin{eqnarray*}
S_{p,n} (v_{p,n} ) (x) & = & \frac{Q_{p} (v
)(x)}{\lambda_{p-1}h_{p-1,n}(x)}\llVert
h_{p,n}\rrVert_{v}
\\
& \leq& \frac{v(x)}{\lambda_{p-1}h_{p-1,n}(x)}e^{-\delta d}\llVert
h_{p,n}\rrVert
_{v}
\\
& = & v_{p,n-1}(x)\frac{e^{-\delta d}}{\lambda_{p-1}}\frac{\llVert
h_{p,n}\rrVert_{v}}{\llVert h_{p-1,n}\rrVert_{v}},
\end{eqnarray*}
where \ref{hypdrift} has been applied.

For $x\in C_{d}$, from Lemma \ref{lemeigfuncmeas} and
\ref{hypminoronly},
\[
\lambda_{p-1}h_{p-1,n}(x)=Q_{p} (h_{p,n}
) (x)\geq\varepsilon_{d}^{-}\nu_{d} (
\mathbb{I}_{C_{d}}h_{p,n} )
\]
and thus using \ref{hypdrift},
\begin{eqnarray*}
S_{p,n} (v_{p,n} ) (x) & \leq& e^{d(1-\delta)+b_{d}}
\frac{\llVert h_{p,n}\rrVert_{v}}{\lambda_{p-1}}\frac{1}{h_{p-1,n}(x)}
\\
& \leq& \frac{e^{d(1-\delta)+b_{d}}}{\varepsilon_{d}^{-}}\llVert
h_{p,n}\rrVert_{v}
\frac{1}{\nu_{d} (\mathbb
{I}_{C_{d}}h_{p,n} )}.
\end{eqnarray*}
We have $\rho_{p,n}<\infty$ and $B_{p,n}<\infty$ because for any
$p\leq n$, $\lambda_{p-1}>0$, $h_{p,n}\in\mathcal{L}_{v}$, $h_{p,n}(x)>0$
for all $x\in\mathsf{X}$, and for any $d\geq\underline{d}$, $\nu
_{d} (C_{d} )>0$.
\end{pf}
%
%le6 #&#
\begin{lem}
\label{lemdriftbound}Assume \ref{hypdrift}--\ref{hypminoronly}.
Then for any $d\in[\underline{d},\infty)$, $0\leq p<q\leq n$
and $x\in\mathsf{X}$,
%
%e33 #&#
\begin{eqnarray}\label{eqtwisteddriftpartial}\quad
& & \check{\mathbb{E}}_{p,x}^{ (n )} \bigl[v_{q,n} (
\check{X}_{q,n} ) \bigr]
\nonumber
\\
& &\qquad \leq\frac{e^{-\delta d (q-p )}}{\prod_{k=p}^{q-1}\lambda
_{k}}\frac{\llVert h_{q,n}\rrVert_{v}}{\llVert h_{p,n}\rrVert
_{v}}v_{p,n} (x )
\\
& &\qquad\quad{} +\frac{e^{d(1-\delta)+b_{d}}}{\varepsilon_{d}^{-}}\llVert
h_{q,n}\rrVert_{v} \Biggl[
\frac{1}{\nu_{d} (\mathbb
{I}_{C_{d}}h_{q,n} )}+\sum_{k=p+1}^{q-1}
\frac{e^{-\delta d
(q-k )}}{\prod_{j=k}^{q-1}\lambda_{j}}\frac{1}{\nu_{d} (\mathbb
{I}_{C_{d}}h_{k,n} )} \Biggr]\nonumber
\end{eqnarray}
with the convention that the sum is zero when $p=q-1$.
\end{lem}
\begin{pf}
For each $n$, $p$ and $q$ in the specified ranges, the proof begins
by recursive application of the drift inequalities of Lemma \ref{lemSdrift}.
A simple induction yields
%
%e34 #&#
\begin{eqnarray}\label{eqEvrecurse}
& & \check{\mathbb{E}}_{p,x}^{ (n )} \bigl[v_{q,n} (
\check{X}_{q,n} ) \bigr]
\nonumber\\[-8pt]\\[-8pt]
& &\qquad \leq\Biggl(\prod_{k=p+1}^{q}
\rho_{k,n} \Biggr)v_{p,n} (x )+\sum
_{k=p+1}^{q} \Biggl(\prod
_{j=k+1}^{q}\rho_{j,n} \Biggr)B_{k,n}\nonumber
\end{eqnarray}
with the convention that the right-most product is equal to $1$
when $p=q-1$.

By the definitions of $ (h_{p,n} )$, $ (\rho_{p,n} )$
and $ (B_{p,n} )$,
%
%e35 #&#
\begin{eqnarray}\label{eqprodofrho}
\prod_{k=p+1}^{q}\rho_{k,n} & = &
\prod_{k=p+1}^{q}\frac{e^{-\delta
d}}{\lambda_{k-1}}
\frac{\llVert h_{k,n}\rrVert_{v}}{\llVert
h_{k-1,n}\rrVert_{v}}
\nonumber\\[-8pt]\\[-8pt]
& = & \frac{e^{-\delta d (q-p )}}{\prod_{k=p+1}^{q}\lambda
_{k-1}}\frac{\llVert h_{q,n}\rrVert_{v}}{\llVert
h_{p,n}\rrVert_{v}}\nonumber
\end{eqnarray}
and for $k<q$,
%
%e36 #&#
\begin{eqnarray}\label{eqprodrhoB}
& & \Biggl(\prod_{j=k+1}^{q}
\rho_{j,n} \Biggr)B_{k,n}
\nonumber
\\
& &\qquad = \biggl(\frac{e^{-\delta d (q-k )}}{\prod_{j=k+1}^{q}\lambda
_{j-1}}\frac{\llVert h_{q,n}\rrVert
_{v}}{\llVert h_{k,n}\rrVert_{v}} \biggr)\frac{e^{d(1-\delta
)+b_{d}}}{\varepsilon_{d}^{-}}
\llVert h_{k,n}\rrVert_{v}\frac
{1}{\nu_{d} (\mathbb{I}_{C_{d}}h_{k,n} )}
\\
& &\qquad =\frac{e^{d(1-\delta)+b_{d}}}{\varepsilon_{d}^{-}}\llVert
h_{q,n}\rrVert_{v}
\frac{e^{-\delta d (q-k )}}{\prod_{j=k}^{q-1}\lambda_{j}}\frac{1}{\nu
_{d} (\mathbb
{I}_{C_{d}}h_{k,n} )}.\nonumber
\end{eqnarray}
The proof is complete upon combining (\ref{eqEvrecurse}), (\ref
{eqprodofrho}),
(\ref{eqprodrhoB}) and applying the definition of $B_{q,n}$ for
the case $q=k$.
\end{pf}

\begin{pf*}{Proof of Proposition \ref{propetaboundedinv}}
For $n=0$ we have trivially $\eta_{0} (v )=\mu(v )$.

For $n\geq1$, by Lemma \ref{lemchangeofmeasure},
%
%e37 #&#
\begin{eqnarray}\label{eqetanvchangeofmeas}
\eta_{n} (v ) & = & \frac{\mathbb{E}_{\mu} [\prod_{q=0}^{n-1}G_{q}
(X_{q} )v (X_{n} ) ]}{\mathbb
{E}_{\mu} [\prod_{q=0}^{n-1}G_{q} (X_{q} )
]}
\nonumber
\\
& = & \int\mu(dx )h_{0,n}(x)\check{\mathbb{E}}_{x}^{(n)}
\bigl[v (\check{X}_{n,n} ) \bigr]
\\
& \leq& \int\mu(dx )h_{0,n}(x)\check{\mathbb{E}}_{x}^{(n)}
\bigl[v_{n,n} (\check{X}_{n,n} ) \bigr],\nonumber
\end{eqnarray}
where the inequality is due to $h_{n,n}=1$ and $\llVert h_{n,n}\rrVert
_{v}\leq1$.
The proof proceeds by bounding the expectation.

Fix $d\in[\underline{d},\infty)$ arbitrarily. Applying
Lemma \ref{lemdriftbound} with $q=n$ and $p=0$, and again noting
$h_{n,n}=1$, $\llVert h_{n,n}\rrVert_{v}\leq1$, we obtain
%
%e38 #&#
\begin{eqnarray}\label{eqEVnnbound}
\check{\mathbb{E}}_{x}^{ (n )} \bigl[v_{n,n} (
\check{X}_{n,n} ) \bigr] & \leq& \frac{e^{-\delta dn}}{\prod
_{k=0}^{n-1}\lambda_{k}}
\frac{1}{\llVert h_{0,n}\rrVert
_{v}}v_{0,n} (x )
\nonumber
\\
& &{} +\frac{e^{d(1-\delta)+b_{d}}}{\varepsilon_{d}^{-}} \Biggl[\frac{1}{\nu
_{d} (C_{d} )}+\sum
_{k=1}^{n-1}\frac{e^{-\delta d
(n-k )}}{\prod_{j=k}^{n-1}\lambda_{j}}\frac{1}{\nu_{d} (\mathbb
{I}_{C_{d}}h_{k,n} )}
\Biggr]
\nonumber\\[-8pt]\\[-8pt]
& = & \frac{e^{-\delta dn}}{\mu Q_{0,n}(1)}\frac{1}{\llVert
h_{0,n}\rrVert_{v}}v_{0,n} (x )
\nonumber
\\
& &{} +\frac{e^{d(1-\delta)+b_{d}}}{\varepsilon_{d}^{-}} \Biggl[\frac{1}{\nu
_{d} (C_{d} )}+\sum
_{k=1}^{n-1}\frac{e^{-\delta d
(n-k )}}{\nu_{d} [\mathbb{I}_{C_{d}}Q_{k,n} (1 )
]} \Biggr]\nonumber
\end{eqnarray}
with the convention (as per Lemma \ref{lemdriftbound}), that the
summation is equal to zero when $n=1$. The equality is due to the
definitions of the $\lambda$-values and $h$-functions.

We now obtain lower bounds in order to treat the $\mu Q_{0,n}(1)$
and\break $\nu_{d} [\mathbb{I}_{C_{d}}Q_{k,n} (1
) ]$
terms. Recall that $d\in[\underline{d},\infty)$ was arbitrary.
Now choose arbitrarily $r\in[\underline{d},d ]$. Then under
\ref{hypminoronly}, for any $\eta\in\mathcal{P}_{v}$ and any
$0\leq k<n$,
%
%e39 #&#
\begin{eqnarray}\label{eqetaQknlowerbound}
\eta\bigl[\mathbb{I}_{C_{d}}Q_{k,n} (1 ) \bigr] & = & \mathbb
{E}_{k,\eta} \Biggl[\mathbb{I}_{C_{d}} (X_{k} )\prod
_{q=k}^{n-1}G_{q}
(X_{q} ) \Biggr]
\nonumber
\\[-2pt]
& \geq& \mathbb{E}_{k,\eta} \Biggl[\mathbb{I}_{C_{r}}
(X_{k} )\prod_{q=k}^{n-1}G_{q}
(X_{q} )\mathbb{I}_{C_{r}} (X_{q} )
\mathbb{I}_{C_{r}} (X_{n} ) \Biggr]
\\[-2pt]
& \geq& \eta(C_{r} ) \bigl[\varepsilon_{r}^{-}
\nu_{r} (C_{r} ) \bigr]^{n-k}.\nonumber
\end{eqnarray}
Under \ref{hypmuv}, for $r$ and $d$ increased if necessary,
but still subject to $r\leq d$, we have $\mu(C_{r} )=1-\mu
(C_{r}^{c} )\geq1-\mu(\mathbb{I}_{C_{r}^{c}}e^{V}
)e^{-r}\geq1-\mu(e^{V} )e^{-r}>0$.
Now hold $r$ constant and if necessary, increase $d$ so that
$e^{-\delta d}< [\varepsilon_{r}^{-}\nu_{r} (C_{r} ) ]^{-1}$.
Equation (\ref{eqetaQknlowerbound}) then gives
%
%e40 #&#
\begin{equation}\label{eqvboundedlead}
\sup_{n\geq1}\frac{e^{-\delta dn}}{\mu Q_{0,n}(1)}\leq\sup_{n\geq1}
\frac
{e^{-\delta dn}}{\mu[\mathbb{I}_{C_{d}}Q_{0,n}(1) ]}\leq\frac
{1}{\mu(C_{r} )}<\infty.
\end{equation}
Then under \ref{hypdrift}, noting $\nu_{d} (C_{r} )>0$
and applying (\ref{eqetaQknlowerbound}),
%
%e41 #&#
\begin{eqnarray}\label{eqvboundedtail}
& & \sup_{n\geq1} \Biggl[\frac{1}{\nu_{d} (C_{d} )}+\sum
_{k=1}^{n-1}\frac{e^{-\delta d (n-k )}}{\nu_{d} [\mathbb
{I}_{C_{d}}Q_{k,n} (1 ) ]} \Biggr]
\nonumber\\[-8pt]\\[-8pt]
& &\qquad \leq\frac{1}{\nu_{d} (C_{d} )}+\frac{1}{\nu_{d}
(C_{r} )}\sup_{n\geq1}
\Biggl[\sum_{k=1}^{n-1}\frac{e^{-\delta d
(n-k )}}{ [\varepsilon_{r}^{-}\nu_{r} (C_{r} )
]^{ (n-k )}}
\Biggr]<\infty.\nonumber
\end{eqnarray}
Combining (\ref{eqvboundedlead}), (\ref{eqvboundedtail}) and
(\ref{eqEVnnbound}), establishes that there exists a finite
constant $c_{\mu}$, independent of $n$ such that
\[
\check{\mathbb{E}}_{x}^{ (n )} \bigl[v_{n,n} (
\check{X}_{n,n} ) \bigr] \leq \frac{1}{\mu(C_{r} )}
\frac
{1}{\llVert h_{0,n}\rrVert_{v}}v_{0,n} (x )+c_{\mu},
\]
and then returning to (\ref{eqetanvchangeofmeas}), we have
shown that
\begin{eqnarray*}
\eta_{n} (v ) & \leq& \frac{1}{\mu(C_{r} )}\frac
{1}{\llVert h_{0,n}\rrVert_{v}}\int
h_{0,n}(x)v_{0,n} (x )\mu(d\mathrmm{x} )
\\[-2pt]
& &{} +c_{\mu}\int h_{0,n}(x)\mu(d\mathrmm{x} )
\\[-2pt]
& = & \frac{\mu(v )}{\mu(C_{r} )}+c_{\mu},
\end{eqnarray*}
where the final equality uses the definition of $v_{0,n}$ and the
property $\mu(h_{0,n} )=\eta(h_{0,n} )=1$ as
in Lemma \ref{lemeigfuncmeas}. Thus there exists a finite constant
$c_{\mu}'$ such that
\[
\sup_{n\geq1}\eta_{n} (v )\leq c_{\mu}'
\mu(v ),
\]
which completes the proof.
\end{pf*}\eject
%
%le7 #&#
\begin{lem}
\label{lemtightness12}Assume \ref{hypdrift}--\ref{hypminoronly}
and let $v$ be as therein. Then
%
%e42 #&#
\begin{equation}\label{eqtightness}
\sup_{n\geq0}\llVert\eta_{n}\rrVert
_{v}<\infty \quad\Longleftrightarrow\quad \inf_{n\geq0}
\lambda_{n}>0.
\end{equation}
\end{lem}
\begin{pf}
$(\Rightarrow)$. Suppose $\sup_{n\geq0}\llVert\eta_{n}\rrVert
_{v}<\infty$.
Then there exists a finite constant $\bar{\eta}$ such that for any
$d\geq\underline{d}$,
\[
\sup_{n\geq0}\eta_{n} \bigl(C_{d}^{c}
\bigr)\leq\sup_{n\geq0}\frac{\eta
_{n} (\mathbb{I}_{C_{d}^{c}}e^{V} )}{e^{d}}\leq\sup
_{n\geq
0}\frac{\eta_{n} (e^{V} )}{e^{d}}\leq\bar{\eta}e^{-d}.
\]
Thus for all $\beta<1$, there exists $d\geq\underline{d}$ such that
$\sup_{n\geq0}\eta_{n} (C_{d}^{c} )<\beta$. Thus for $\beta\in(0,1)$
there exists $r\geq\underline{d}$ such that
\begin{eqnarray*}
\inf_{n\geq0}\lambda_{n}&\geq&\inf_{n\geq0}
\eta_{n} \bigl(\mathbb{I}_{C_{r}}Q_{n+1} (
\mathbb{I}_{C_{r}} ) \bigr)\geq\varepsilon_{r}^{-}
\nu_{r} (C_{r} )\inf_{n\geq0}
\eta_{n} (C_{r} )\\
&\geq&\varepsilon_{r}^{-}
\nu_{r} (C_{r} ) (1-\beta),
\end{eqnarray*}
where the second inequality is due to \ref{hypminoronly}.

$(\Leftarrow)$. Suppose $\inf_{n\geq0}\lambda_{n}>0$. Then there
exists $\underline{\lambda}>0$ such that for any $n\geq1$,
\[
\eta_{n} \bigl(e^{V} \bigr)=\frac{\eta_{n-1}Q_{n} (e^{V} )}{\eta
_{n-1} (G_{n-1} )}\leq
\frac{\eta_{n-1}Q_{n} (e^{V}
)}{\underline{\lambda}},
\]
where (\ref{eqeigfuncmeas}) has been used. Now set $d>\underline
{d}\vee(-\frac{1}{\delta}\log\underline{\lambda} )$.
Then under \ref{hypdrift},
%
%e43 #&#
\begin{eqnarray}\label{eqetanVrecurse}
\eta_{n} \bigl(e^{V} \bigr) & \le& \frac{\eta_{n-1} [\mathbb
{I}_{C_{d}^{c}}Q_{n} (e^{V} ) ]}{\underline{\lambda
}}+
\frac{\eta_{n-1} [\mathbb{I}_{C_{d}}Q_{n} (e^{V} )
]}{\underline{\lambda}}
\nonumber
\\
& \leq& \frac{e^{-\delta d}}{\underline{\lambda}}\eta_{n-1} \bigl(e^{V}
\bigr)+
\frac{e^{d(1-\delta)+b_{d}}}{\underline{\lambda
}}
\\
& = :& \rho\eta_{n-1} \bigl(e^{V} \bigr)+B\nonumber
\end{eqnarray}
for some $\rho<1$ and $B<\infty$. Iteration of (\ref{eqetanVrecurse})
establishes $(1)$.
\end{pf}
%
%le8 #&#
\begin{lem}
\label{lemtightness2to3}Assume \ref{hypdrift}--\ref{hypminorandmaj}
and let $v$ be as therein. Then
\[
\inf_{n\geq0}\lambda_{n}>0 \quad\Longrightarrow\quad\sup
_{n\geq
1}\sup_{0\leq p\leq n}\llVert h_{p,n}
\rrVert_{v}<\infty.
\]
\end{lem}
\begin{pf}
Recall the definition
%
%e44 #&#
\begin{equation}\label{eqhnpreminder}
h_{p,n}(x)=\frac{Q_{p.n} (1 )(x)}{\eta_{p}Q_{p,n} (1)}.
\end{equation}
For the case $p=n$, $h_{p,n}=1$. For other cases we proceed by decomposing
and then bounding the numerator.

Set $d\in[\underline{d},\infty)$ arbitrarily, let $n\geq1$,
$0\leq p<n$ and define $\tau_{p}^{(d)}:=\inf\{ q\geq p;X_{q}\in
C_{d},X_{q+1}\in C_{d} \} $.
Now consider the decomposition
%
%e45 #&#
\begin{eqnarray}\label{eqfirstentrancedecomp}
Q_{p,n}(1) (x) & = & \sum_{k=p}^{n-1}
\mathbb{E}_{p,x} \Biggl[\prod_{q=p}^{n-1}G_{q}
(X_{q} )\mathbb{I} \bigl\{ \tau_{p}^{(d)}=k
\bigr\} \Biggr]
\nonumber\\[-8pt]\\[-8pt]
& &{} +\mathbb{E}_{p,x} \Biggl[\prod_{q=p}^{n-1}G_{q}
(X_{q} )\mathbb{I} \bigl\{ \tau_{p}^{(d)}\geq n
\bigr\} \Biggr]\nonumber
\end{eqnarray}
and define
\begin{eqnarray*}
A_{p}&:=&\interleave\mathbb{I}_{C_{d}^{c}}Q_{p}
\interleave_{v},\qquad B_{p}:=\interleave\mathbb{I}_{C_{d}}Q_{p}
\interleave_{v},
\qquad
\Xi_{0}:=v (X_{p} ),\\
\Xi_{j}&:=& \Biggl[\prod
_{q=p}^{p+j-1}\frac{G_{q} (X_{q} )}{A_{q+1}^{\mathbb
{I}_{C_{d}^{c}} (X_{q} )}B_{q+1}^{\mathbb{I}_{C_{d}}
(X_{q} )}} \Biggr]v
(X_{p+j} ),\qquad 1\leq j\leq n-p.
\end{eqnarray*}
Assumption \ref{hypdrift} implies that, for $1\leq j\leq n-p$, $\mathbb
{E}_{p+j-1,X_{p+j-1}} [\Xi_{j} ]\leq\Xi_{j-1}$,
so that
%
%e46 #&#
\begin{equation}\label{eqfilsupermart-1}
\mathbb{E}_{p,x} [\Xi_{n-p} ] \leq \mathbb{E}_{p,x}
[\Xi_{0} ]=v(x).
\end{equation}
For $k>p$, define $M_{p,k}^{(d)}:=\sum_{q=p}^{k-1}\mathbb
{I}_{C_{d}^{c}} (X_{q} )$.
Then the following bound holds under \ref{hypdrift}:
%
%e47 #&#
\begin{eqnarray}\label{eqABunivariate}\quad
& & \Biggl[\prod_{q=p}^{k-1}A_{q+1}^{\mathbb{I}_{C_{d}^{c}}
(X_{q} )}B_{q+1}^{\mathbb{I}_{C_{d}} (X_{q} )}
\Biggr]\mathbb{I} \bigl\{ M_{p,k}^{(d)}\geq(k-p )/2 \bigr\}
\nonumber
\\
& &\qquad \leq\biggl(\sup_{q\geq1}\interleave\mathbb
{I}_{C_{d}^{c}}Q_{q}\interleave_{v}
\biggr)^{M_{p,k}^{(d)}}\mathbb{I} \bigl\{ M_{p,k}^{(d)}\geq
(k-p )/2 \bigr\} \biggl(1\vee\sup_{q\geq1}\interleave
Q_{q}\interleave_{v} \biggr)^{ (k-p
)/2}
\\
& &\qquad \leq\exp\bigl[-\delta d (k-p )/2 \bigr]\exp\bigl[b_{\underline
{d}} (k-p
)/2 \bigr],\nonumber
\end{eqnarray}
where $\interleave\mathbb{I}_{C_{d}} Q_{q}\interleave_{v}\leq
\interleave Q_{q}\interleave_{v}$
has been used.

Consider one term from the summation in (\ref{eqfirstentrancedecomp})
with $p<k<n$. By Douc et al. [(\citeyear{filtertheDFMP09}), Lemma 17]
\[
\mathbb{I} \bigl\{ \tau_{p}^{(d)}\geq k \bigr\} =\mathbb
{I} \Biggl\{ \sum_{q=p}^{k-1}
\mathbb{I}_{C_{d}} (X_{q} )\mathbb{I}_{C_{d}}
(X_{q+1} )=0 \Biggr\} \leq\mathbb{I} \bigl\{ M_{p,k}^{(d)}
\geq(k-p )/2 \bigr\}.
\]
Then combining (\ref{eqfilsupermart-1}) and (\ref{eqABunivariate})
and using \ref{hypminorandmaj},
%
%e48 #&#
\begin{eqnarray}\label{eqhabovekgreatp}
& & \mathbb{E}_{p,x} \Biggl[\prod_{q=p}^{n-1}G_{q}
(X_{q} )\mathbb{I} \bigl\{ \tau_{p}^{(d)}=k \bigr
\} \Biggr]
\nonumber
\\
& &\qquad \leq\varepsilon_{d}^{+}\nu_{d} \bigl[\mathbb
{I}_{C_{d}}Q_{k+1,n}(1) \bigr]\mathbb{E}_{p,x} \Biggl[
\prod_{q=p}^{k-1}G_{q}
(X_{q} )\mathbb{I} \bigl\{ M_{p,k}^{(d)}\geq(k-p
)/2 \bigr\} v (X_{k} ) \Biggr]
\nonumber\\[-8pt]\\[-8pt]
& &\qquad \leq\varepsilon_{d}^{+}\nu_{d} \bigl[\mathbb
{I}_{C_{d}}Q_{k+1,n}(1) \bigr]v(x)
\nonumber
\\
& &\qquad\quad{} \times\exp\bigl[-\delta d (k-p )/2 \bigr]\exp\bigl[b_{\underline
{d}} (k-p
)/2 \bigr],\qquad k>p,\nonumber
\end{eqnarray}
and similarly,
%
%e49 #&#
\begin{eqnarray}\label{eqhabovekeqn}
& & \mathbb{E}_{p,x} \Biggl[\prod_{q=p}^{n-1}G_{q}
(X_{q} )\mathbb{I} \bigl\{\tau_{d}^{(p)}\geq n
\bigr\} \Biggr]
\nonumber
\\
& &\qquad \leq\mathbb{E}_{p,x} \Biggl[\prod_{q=p}^{n-1}G_{q}
(X_{q} )\mathbb{I} \bigl\{ M_{p,k}^{(d)}\geq(n-p
)/2 \bigr\} v (X_{n} ) \Biggr]
\\
& &\qquad \leq v(x)\exp\bigl[-\delta d (n-p )/2 \bigr]\exp\bigl
[b_{\underline{d}}
(n-p )/2 \bigr]\nonumber
\end{eqnarray}
and also by \ref{hypminorandmaj},
%
%e50 #&#
\begin{equation}\label{eqhabovekeqp}
\mathbb{E}_{p,x} \Biggl[\prod_{q=p}^{n-1}G_{q}
(X_{q} )\mathbb{I} \bigl\{\tau_{d}^{(p)}=p \bigr\}
\Biggr] \leq \varepsilon_{d}^{+}\nu_{d} \bigl[
\mathbb{I}_{C_{d}}Q_{p+1,n}(1) \bigr]v(x),
\end{equation}
recalling from Section \ref{subFKsetup} the convention $Q_{n+1,n}=\mathit{Id}$.
Returning to (\ref{eqfirstentrancedecomp}), the bounds of (\ref
{eqhabovekgreatp})--(\ref{eqhabovekeqp})
show that for $p<n$,
%
%e51 #&#
\begin{eqnarray}\label{eqQpnboundabove}
& & Q_{p,n}(1) (x)
\nonumber
\\
& &\qquad \leq\varepsilon_{d}^{+}v(x)\sum
_{k=p}^{n-1}\exp\bigl[-\delta d (k-p )/2 \bigr]\exp
\bigl[b_{\underline{d}} (k-p )/2 \bigr]\nu_{d} \bigl[
\mathbb{I}_{C_{d}}Q_{k+1,n}(1) \bigr]
\\
& &\qquad\quad{} +v(x)\exp\bigl[-\delta d (n-p )/2 \bigr]\exp\bigl[b_{\underline
{d}} (n-p
)/2 \bigr].\nonumber
\end{eqnarray}
We now turn to the denominator of (\ref{eqhnpreminder}) and stress
that we are continuing to use the same arbitrary value of $d$ as
above.

As per the statement of the lemma, suppose $\underline{\lambda}:=\inf
_{n\geq0}\lambda_{n}>0$.
Then by Lem\-ma~\ref{lemtightness12}, $\bar{\eta}:=\sup_{n\geq0}\eta
_{n} (e^{V} )<\infty$
and choosing independently $\varepsilon\in(0,1)$, by \ref{hypdrift}
$d$ may then be chosen large enough that
\[
\inf_{n\geq0}\eta_{n} (C_{d} )=\inf
_{n\geq0}1-\eta_{n} \bigl(C_{d}^{c}
\bigr)\geq\inf_{n\geq0}1-\eta_{n} \bigl(\mathbb
{I}_{C_{d}^{c}}e^{V} \bigr)e^{-d}\geq1-\bar{
\eta}e^{-d}\geq1-\varepsilon=:\underline{\eta}.
\]
Then for $p<k<n$,
%
%e52 #&#
\begin{eqnarray}\label{eqetaQpnlowerbound}
\eta_{p}Q_{p,n}(1) & = & \Biggl(\prod
_{q=p}^{k-1}\lambda_{q} \Biggr)\eta
_{k}Q_{k,n}(1)
\nonumber
\\
& \geq& \underline{\lambda}^{ (k-p )}\eta_{k} \bigl[\mathbb
{I}_{C_{d}}Q_{k,n}(1) \bigr]
\nonumber\\[-8pt]\\[-8pt]
& \geq& \varepsilon_{d}^{-}\underline{\lambda}^{ (k-p )}
\eta_{k} (C_{d} )\nu_{d} \bigl[
\mathbb{I}_{C_{d}}Q_{k+1,n}(1) \bigr]
\nonumber
\\
& \geq& \varepsilon_{d}^{-}\underline{\lambda}^{ (k-p
)}
\underline{\eta}\nu_{d} \bigl[\mathbb{I}_{C_{d}}Q_{k+1,n}(1)
\bigr]\nonumber
\end{eqnarray}
and for $p<n$,
%
%e53 #&#
\begin{equation}\label{eqetaQpnlowerbound2}
\eta_{p}Q_{p,n}(1) \geq \varepsilon_{d}^{-}
\eta_{p} (C_{d} )\nu_{d} \bigl[
\mathbb{I}_{C_{d}}Q_{p+1,n}(1) \bigr]\geq\varepsilon
_{d}^{-}\underline{\eta}\nu_{d} \bigl[\mathbb
{I}_{C_{d}}Q_{p+1,n}(1) \bigr]
\end{equation}
and also
%
%e54 #&#
\begin{equation}\label{eqetaQpnlowerbound3}
\eta_{p}Q_{p,n}(1)\geq\underline{\lambda}^{ (n-p )}.
\end{equation}
Combining (\ref{eqetaQpnlowerbound})--(\ref{eqetaQpnlowerbound3})
with (\ref{eqQpnboundabove}) and (\ref{eqhnpreminder}),
we finally obtain, for $p<n$,
\begin{eqnarray*}
h_{p,n}(x) & \leq& \frac{\varepsilon_{d}^{+}}{\varepsilon_{d}^{-}\underline
{\eta}}v(x)\sum
_{k=p}^{n-1}\exp\bigl[-\delta d (k-p )/2 \bigr]\exp
\bigl[ (k-p ) (b_{\underline{d}}/2-\log\underline{\lambda} ) \bigr]
\\
&&{} + v(x)\exp\bigl[-\delta d (n-p )/2 \bigr]\exp\bigl[ (n-p )
(b_{\underline{d}}/2-\log\underline{\lambda} ) \bigr].
\end{eqnarray*}
Then increasing $d$ further if necessary, we conclude that there
exists $c<\infty$ such that for any $x\in\mathsf{X}$, $\sup_{n\geq1}\sup
_{0\leq p\le n}h_{p,n}(x)\leq cv(x)$,
and this completes the proof.
\end{pf}
%
%le9 #&#
\begin{lem}
\label{lemhpnboundedinv}Assume \ref{hypdrift}--\ref{hypminoronly},
and let $v$ be as therein. Then
\[
\sup_{n\geq1}\sup_{0\leq p\leq n}\llVert
h_{p,n}\rrVert_{v}<\infty\quad\Longrightarrow\quad\inf
_{n\geq0}\lambda_{n}>0.
\]
\end{lem}
\begin{pf}
Suppose $\sup_{n\geq1}\sup_{0\leq p\leq n}\llVert h_{p,n}\rrVert
_{v}<\infty$.
Then by Lemma \ref{lemeigfuncmeas} and\break \ref{hypminoronly},
for any $x\in C_{d}$,
\begin{eqnarray*}
\inf_{n\geq0}\lambda_{n} & = & \inf
_{n\geq0}\frac{Q_{n+1}
(h_{n+1,n+1} )(x)}{h_{n,n+1}(x)}
\\
& \geq& \inf_{n\geq0}\frac{Q_{n+1} (C_{d} )(x)}{\llVert
h_{n,n+1}\rrVert_{v}v(x)}
\\
& \geq& \frac{\varepsilon_{d}^{-}}{e^{d}}\frac{\nu_{d} (C_{d}
)}{\sup_{n\geq0}\llVert h_{n,n+1}\rrVert_{v}}
\\
& > & 0.
\end{eqnarray*}
\upqed\end{pf}

\subsection*{Proofs for Section \protect\ref{subMET}}

The following lemma will be used in the proofs of Theorems \ref{thmMET}
and \ref{thmexpmoments}.
%
%le10 #&#
\begin{lem}
\label{lemhpnboundedbelowonC} Assume \ref{hypdrift}--\ref
{hypminorandmaj},
and let $\underline{d}$ be as therein. Then for any $d\in
[\underline{d},\infty)$,
\[
\inf_{n\geq1}\inf_{0\leq p\leq n}\inf
_{x\in C_{d}}h_{p,n} (x )>0.
\]
\end{lem}
\begin{pf}
We will prove a finite, uniform upper bound on
%
%e55 #&#
\begin{eqnarray}\label{eq1onhpn}
\sup_{x\in C_{d}}\frac{1}{h_{p,n}(x)} & = & \sup_{x\in C_{d}}
\frac{\eta
_{p}Q_{p,n} (1 )}{Q_{p,n}(1)(x)}
\nonumber\\[-8pt]\\[-8pt]
& = & \frac{\eta_{p}Q_{p,n} (1 )}{\inf_{x\in
C_{d}}Q_{p,n}(1)(x)}.\nonumber
\end{eqnarray}
The proof uses the same approach as in the proof of Lemma \ref
{lemhpnboundedinv},
and therefore some steps are omitted for brevity. For the case $p=n$,
$\eta_{n}Q_{n,n}(1)=1$ and \mbox{$Q_{n,n}(1)(x)=1$} for all $x$. For the
remaining cases we proceed by considering the numerator of
(\ref{eq1onhpn}).

Set $d\in[\underline{d},\infty)$ arbitrarily, let $n\geq1$
and $p<n$ and define $\tau_{p}^{(d)}:=\inf\{ q\geq p;X_{q}\in
C_{d},X_{q+1}\in C_{d} \} $.
We have the decomposition
%
%e56 #&#
\begin{eqnarray}\label{eqetaQpndecomp}
\eta_{p}Q_{p,n} (1 ) & = & \sum
_{k=p}^{n-1}\mathbb{E}_{p,\eta
_{p}} \Biggl[\prod
_{q=p}^{n-1}G_{q}
(X_{q} )\mathbb{I} \bigl[\tau_{p}^{(d)}=k \bigr]
\Biggr]
\nonumber\\[-8pt]\\[-8pt]
& &{} +\mathbb{E}_{p,\eta_{p}} \Biggl[\prod_{q=p}^{n-1}G_{q}
(X_{q} )\mathbb{I} \bigl[\tau_{p}^{(d)}\geq n
\bigr] \Biggr].\nonumber
\end{eqnarray}

This is of exactly the same form as in equation (\ref{eqfirstentrancedecomp})
in the proof of Lemma \ref{lemhpnboundedinv}, except for the
initial measure $\eta_{p}$. Thus by exactly the same arguments [integrate
equation (\ref{eqQpnboundabove}) w.r.t. $\eta_{p}$] we obtain
the bound
%
%e57 #&#
\begin{eqnarray}\label{eqetapQpnbound}
& & \eta_{p}Q_{p,n} (1 )
\nonumber
\\
& &\qquad \leq\varepsilon_{d}^{+}\eta_{p} (v )\sum
_{k=p}^{n-1}\exp\bigl[-\delta d (k-p )/2
\bigr]\exp\bigl[b_{\underline{d}} (k-p )/2 \bigr]\nu_{d} \bigl[
\mathbb{I}_{C_{d}}Q_{k+1,n}(1) \bigr]
\\
& &\qquad\quad{} +\eta_{p} (v )\exp\bigl[-\delta d (n-p )/2 \bigr]\exp
\bigl[b_{\underline{d}} (n-p )/2 \bigr].\nonumber
\end{eqnarray}
Now set $r\in[\underline{d},d ]$. For the denominator of
(\ref{eq1onhpn}) we have by \ref{hypminoronly},
%
%e58 #&#
\begin{eqnarray}\label{eqinfQpn}
\inf_{x\in C_{d}}Q_{p,n}(1) (x) & \geq& \inf
_{x\in C_{d}}Q_{p} \bigl[\mathbb{I}_{C_{d}}Q_{p+1,n}(1)
\bigr](x)
\nonumber\\[-8pt]\\[-8pt]
& \geq& \varepsilon_{d}^{-}\nu_{d} \bigl[\mathbb
{I}_{C_{r}}Q_{p+1,n}(1) \bigr],\nonumber
\end{eqnarray}
also
\[
\varepsilon_{d}^{-}\nu_{d} \bigl[
\mathbb{I}_{C_{r}}Q_{p+1,n}(1) \bigr] \geq
\varepsilon_{d}^{-}\nu_{d} (C_{r} ) \bigl[
\varepsilon_{r}^{-}\nu_{r} (C_{r} )
\bigr]^{n-p-1}
\]
and for $p<k<n$,
%
%e59 #&#
\begin{eqnarray}\label{eqinfQpn2}
& & \varepsilon_{d}^{-}\nu_{d} \bigl[
\mathbb{I}_{C_{r}}Q_{p+1,n}(1) \bigr]
\nonumber
\\
& &\qquad =\varepsilon_{d}^{-}\mathbb{E}_{p+1,\nu_{d}} \Biggl[
\mathbb{I}_{C_{r}} (X_{p+1} )\prod
_{q=p+1}^{n-1}G_{q} (X_{q} )
\Biggr]
\nonumber
\\
& &\qquad \geq\varepsilon_{d}^{-}\mathbb{E}_{p+1,\nu_{d}} \Biggl[
\mathbb{I}_{C_{r}} (X_{p+1} )\prod
_{q=p+1}^{n-1}G_{q} (X_{q} )
\mathbb{I}_{C_{d}} (X_{k} )\mathbb{I}_{C_{d}}
(X_{k+1} ) \Biggr]
\nonumber\\[-8pt]\\[-8pt]
& &\qquad \geq\varepsilon_{d}^{-}\mathbb{E}_{p+1,\nu_{d}} \Biggl[
\mathbb{I}_{C_{r}} (X_{p+1} )\prod
_{q=p+1}^{k-1}G_{q} (X_{q} )
\Biggr]\varepsilon_{d}^{-}\nu_{d} \bigl[\mathbb
{I}_{C_{d}}Q_{k+1,n}(1) \bigr]
\nonumber
\\
& &\qquad \geq\varepsilon_{d}^{-}\mathbb{E}_{p+1,\nu_{d}} \Biggl[
\mathbb{I}_{C_{r}} (X_{p+1} )\prod
_{q=p+1}^{k-1}G_{q} (X_{q} )
\mathbb{I}_{C_{r}} (X_{q+1} ) \Biggr]\varepsilon_{d}^{-}
\nu_{d} \bigl[\mathbb{I}_{C_{d}}Q_{k+1,n}(1) \bigr]
\nonumber
\\
& &\qquad \geq\varepsilon_{d}^{-}\nu_{d} (C_{r}
) \bigl[\varepsilon_{r}^{-}\nu_{r}
(C_{r} ) \bigr]^{k-p-1}\varepsilon_{d}^{-}\nu
_{d} \bigl[\mathbb{I}_{C_{d}}Q_{k+1,n}(1) \bigr].\nonumber
\end{eqnarray}
Combining (\ref{eq1onhpn}), (\ref{eqetapQpnbound}), (\ref
{eqinfQpn})
and (\ref{eqinfQpn2}) gives for $p<n$,
\begin{eqnarray*}
\sup_{x\in C_{d}}\frac{1}{h_{p,n}(x)}
&\leq&\frac{\varepsilon_{d}^{+}}{\varepsilon_{d}^{-}}\eta_{p} (v )
\\
& &{} +\frac{\varepsilon_{d}^{+}}{\varepsilon_{d}^{-}}\frac{\eta_{p}
(v )}{\varepsilon_{d}^{-}\nu_{d} (C_{r} )}\frac{1}{\varepsilon
_{r}^{-}\nu_{r} (C_{r} )}
\\
& &\hspace*{11pt}{}\times\Biggl(\sum_{k=p+1}^{n-1}\exp
\bigl[- (k-p ) \bigl(\delta d/2-b_{\underline{d}}/2-\log\bigl[
\varepsilon_{r}^{-}\nu_{r} (C_{r} )
\bigr] \bigr) \bigr] \Biggr)
\\
& &{} +\frac{1}{\varepsilon_{d}^{-}}\frac{\eta_{p} (v )}{\nu
_{d} (C_{r} )}\exp\bigl[- (n-p ) \bigl(\delta
d/2-b_{\underline{d}}/2-\log\bigl[\varepsilon_{r}^{-}
\nu_{r} (C_{r} ) \bigr] \bigr) \bigr]
\end{eqnarray*}
with the convention that the summation is zero when $p=n-1$. With
$r$ kept fixed, increasing $d$ and noting that under the assumptions
of the lemma, Proposition \ref{propetaboundedinv} holds, we conclude
that there exists a finite constant $c_{\mu}(d)$ such that
\[
\sup_{n\geq1}\sup_{0\leq p\leq n}\sup
_{x\in C_{d}}\frac
{1}{h_{p,n}(x)}\leq c_{\mu}(d).
\]
The proof is complete because $d_{1}\leq d_{2}\Rightarrow
C_{d_{1}}\subseteq C_{d_{2}}$.
\end{pf}

\begin{pf*}{Proof of Theorem \ref{thmMET}} The proof is based directly
on those
of Douc et al. [(\citeyear{filtertheDFMP09}), Proposition 12 and Lemma 15], which
are in turn developments from the decomposition ideas of \citet{filtertheKV08}.
However, there are some crucial differences here: the focus of the
present work is on the $v$-norm on measures, as opposed to total
variation, and different techniques will be used to deal with and
control denominator terms in equation (\ref{eqfiltstartingdecomp})
below, by way of Propositions \ref{propetaboundedinv} and \ref{propequiv}.

Throughout the proof, $c$ is a finite constant whose value depends
on $\mu$ and the quantities in \ref{hypdrift}--\ref{hypminorandmaj}
and whose value may change on each appearance.

Let $ (\bar{X}_{n};n\geq0 )$ be the bi-variate Markov chain
on $\mathsf{X}^{2}$ with
\[
\bar{X}_{n}| \bigl\{ \bar{X}_{n-1}= \bigl(x_{n-1,}x_{n-1}^{\prime}
\bigr) \bigr\} \sim M_{n} (x_{n-1},\cdot)\otimes
M_{n} \bigl(x_{n-1}^{\prime},\cdot\bigr)
\]
and for some distribution $H$ on $\mathsf{X}^{2}$ we denote by $\bar
{\mathbb{E}}_{H}$
the expectation with respect to the law of this bi-variate chain initialized
by $\bar{X}_{0}\sim H$. In line with previous definitions, for $\eta$
a distribution on $\mathsf{X}$, we write $\bar{\mathbb{E}}_{p,\delta
_{x}\otimes\eta}:=\int\delta_{x} (\mathrmm{d}x )\eta
(\mathrmm{d}x^{\prime} )\bar{\mathbb{E}}_{H}
[\varphi(\bar{X}_{p},\ldots,\bar{X}_{n} )| \{ \bar
{X}_{p}= (x,x^{\prime} ) \} ]$.
Also define $\bar{C}_{d}:=C_{d}\times C_{d}$ and throughout the following
writing $\bar{x}= (x,x^{\prime} )$ for a point in $\mathsf{X}^{2}$,
define $\bar{G}_{n} (\bar{x} ):=G_{n} (x )G_{n}
(x^{\prime} )$
and $\bar{v} (\bar{x} ):=v (x )v (x^{\prime})$.

For each $n\geq1$ define the tensor-product kernel $\bar{Q}_{n}
(\bar{x},d\bar{y} ):=Q_{n}(x,dy)\otimes Q_{n}(x^{\prime},dy^{\prime})$,
and let $ (\bar{Q}_{p,n} )$ be the semigroup defined in
the same fashion as (\ref{eqQsemigroup}). Now fix arbitrarily $d\in
[\underline{d},\infty)$, and define, for $n\geq1$,
\[
\bar{R}_{n} (\bar{x},d\bar{y} ) := \bar{Q}_{n} (\bar
{x},d\bar{y} )-\mathbb{I}_{\bar{C}_{d}} (\bar{x} ) \bigl(\varepsilon_{d}^{-}
\bigr)^{2}\nu_{d}\otimes\nu_{d} (d\bar{y} )
\]
and $ (\bar{R}_{p,n} )$ in the same way. The dependence
of $\bar{R}_{n}$ on $d$ is suppressed from the notation.

First set $n\geq1$ and $0\leq p\leq n$ arbitrarily. We have from
the above definitions,
%
%e60 #&#
\begin{eqnarray}\label{eqfiltstartingdecomp}
& & \biggl\llvert\frac{Q_{p,n} (\varphi)(x)}{h_{p,n}(x)\prod
_{q=p}^{n-1}\lambda_{q}}-\eta_{n} (\varphi)\biggr
\rrvert
\nonumber
\\
& &\qquad =\biggl\llvert\frac{Q_{p,n} (\varphi)(x)}{Q_{p,n} (1
)(x)}-\frac{\eta_{p}Q_{p,n} (\varphi)}{\eta_{p}Q_{p,n}
(1 )}\biggr\rrvert
\nonumber
\\
& &\qquad =\biggl\llvert\frac{ (\delta_{x}\otimes\eta_{p} )\bar
{Q}_{p,n} (\varphi\otimes1 )- (\eta_{p}\otimes\delta
_{x} )\bar{Q}_{p,n} (\varphi\otimes1 )}{Q_{p,n}
(1 )(x)\eta_{p}Q_{p,n} (1 )}\biggr\rrvert
\nonumber\\[-8pt]\\[-8pt]
& &\qquad =\biggl\llvert\frac{ (\delta_{x}\otimes\eta_{p} )\bar
{R}_{p,n} (\varphi\otimes1-1\otimes\varphi)}{Q_{p,n}
(1 )(x)\eta_{p}Q_{p,n} (1 )}\biggr\rrvert
\nonumber
\\
& &\qquad \leq2\llVert\varphi\rrVert_{v}\frac{ (\delta
_{x}\otimes\eta_{p} )\bar{R}_{p,n} (\bar{v}
)}{Q_{p,n} (1 )(x)\eta_{p}Q_{p,n} (1 )}\nonumber\\
&&\qquad=:2\llVert
\varphi\rrVert_{v}\frac{\Delta_{p,n} (x,\eta_{p}
)}{Q_{p,n} (1 )(x)\eta_{p}Q_{p,n} (1 )},\nonumber
\end{eqnarray}
where the third equality is due to the decomposition of \citet{filtertheKV08}.

Now define $\rho_{d}:=1- (\frac{\varepsilon_{d}^{-}}{\varepsilon
_{d}^{+}} )^{2}<1$
and $\bar{M}_{p,n}^{ (d )}:=\sum_{k=p}^{n-1}\mathbb{I}_{\bar
{C}_{d}} (\bar{X}_{k} )\mathbb{I}_{\bar{C}_{d}} (\bar
{X}_{k+1} )$.
Following essentially the same argument as Douc et al. [(\citeyear{filtertheDFMP09}),
proof of Proposition~12],
then gives, for any $\beta\in(0,1)$,
\begin{eqnarray*}
\Delta_{p,n} (x,\eta_{p} ) & \leq& \bar{
\mathbb{E}}_{\delta
_{x}\otimes\eta_{p}} \Biggl[\prod_{q=p}^{n-1}
\bar{G}_{q} (\bar{X}_{q} )\rho_{d}^{\bar{M}_{p,n}^{ (d )}}
\bar{v} (\bar{X}_{n} ) \Biggr]
\\
& = & \bar{\mathbb{E}}_{\delta_{x}\otimes\eta_{p}} \Biggl[\prod
_{q=p}^{n-1}\bar{G}_{q} (
\bar{X}_{q} )\rho_{d}^{\bar
{M}_{p,n}^{ (d )}}\mathbb{I} \bigl\{
\bar{M}_{p,n}^{
(d )}\geq\beta(n-p ) \bigr\} \bar{v} (\bar
{X}_{n} ) \Biggr]
\\
& &{} +\bar{\mathbb{E}}_{\delta_{x}\otimes\eta_{p}} \Biggl[\prod_{q=p}^{n-1}
\bar{G}_{q} (\bar{X}_{q} )\rho_{d}^{\bar
{M}_{p,n}^{ (d )}}
\mathbb{I} \bigl\{ \bar{M}_{p,n}^{
(d )}<\beta(n-p ) \bigr\}
\bar{v} (\bar{X}_{n} ) \Biggr]
\\
& = :& \Delta_{p,n}^{(1)} (x,\eta_{p} )+\Delta
_{p,n}^{(2)} (x,\eta_{p} ).
\end{eqnarray*}
We first consider $\Delta_{p,n}^{(1)} (x,\eta_{p} )$. As
$\rho_{d}<1$, we have the bound
\[
\frac{\Delta_{p,n}^{(1)} (x,\eta_{p} )}{Q_{p,n} (1
)(x)\eta_{p}Q_{p,n} (1 )} \leq \rho_{d}^{\beta
(n-p )} \biggl[
\frac{Q_{p,n}(v)(x)}{Q_{p,n}(1)(x)} \biggr]\eta_{n}(v),
\]
but using Lemma \ref{lemdriftbound}, Propositions \ref{propetaboundedinv},
\ref{propequiv} and Lemma \ref{lemhpnboundedbelowonC}
show that for $r$ large enough, but then fixed,
%
%e61 #&#
\begin{eqnarray}\label{eqQpnvbound}
\frac{Q_{p,n}(v)(x)}{Q_{p,n}(1)(x)} & \leq& \check{\mathbb{E}}_{p,x}^{
(n )}
\bigl[v_{n,n} (\check{X}_{n,n} ) \bigr]
\nonumber
\\
& \leq& \frac{e^{-\delta r (n-p )}}{\underline{\lambda
}^{ (n-p )}}\frac{1}{\llVert h_{p,n}\rrVert
_{v}}v_{p,n} (x )
\nonumber\\[-8pt]\\[-8pt]
& &{} +\frac{e^{r(1-\delta)+b_{r}}}{\varepsilon_{r}^{-}} \Biggl[\frac{1}{\nu
_{r} (\mathbb{I}_{C_{r}}h_{p,n} )}+\sum
_{k=p+1}^{n-1}\frac
{e^{-\delta r (n-k )}}{\underline{\lambda}^{ (n-k
)}}\frac{1}{\nu_{r} (\mathbb{I}_{C_{r}}h_{k,n} )}
\Biggr]
\nonumber
\\
& \leq& c\frac{v_{p,n} (x )}{\llVert h_{p,n}\rrVert
_{v}},\nonumber
\end{eqnarray}
so
%
%e62 #&#
\begin{eqnarray}\label{eqDelta1bound}
\frac{\Delta_{p,n}^{(1)} (x,\eta_{p} )}{Q_{p,n} (1
)(x)\eta_{p}Q_{p,n} (1 )} & \leq& c\rho_{d}^{\beta
(n-p )}
\frac{v_{p,n} (x )}{\llVert h_{p,n}\rrVert
_{v}}\eta_{n}(v)
\nonumber\\[-8pt]\\[-8pt]
& \leq& c\rho_{d}^{\beta(n-p )}\frac{v_{p,n} (x
)}{\llVert h_{p,n}\rrVert_{v}}\mu(v ),\nonumber
\end{eqnarray}
where the second inequality is due to Proposition
\ref{propetaboundedinv}.

Now consider $\Delta_{p,n}^{(2)} (x,\eta_{p} )$. The main
idea for treating this term is that of \citet{filtertheDFMP09}, proof
of Lemma 15.
There are some cosmetic differences of indexing, and some intermediate
steps are omitted for brevity. Define
\begin{eqnarray*}
\widetilde{M}{}^{ (d )}_{p,n}&:=&\sum_{k=p}^{n-1}
\mathbb{I}_{\bar
{C}_{d}^{c}} (\bar{X}_{k} ),\\
a_{p,n}&:=& \bigl
\lfloor(n-p ) (1-\beta)/2-1/2 \bigr\rfloor,
\\
A_{p}&:=&\interleave\mathbb{I}_{\bar{C}_{d}^{c}}\bar{Q}_{p}
\interleave_{v\otimes v},\qquad B_{p}:=\interleave\mathbb{I}_{\bar{C}_{d}}
\bar{Q}_{p}\interleave_{v\otimes v},\qquad
\Xi_{0}:=\bar{v} (\bar{X}_{p} ),\\
\Xi_{k}&:=&
\Biggl[\prod_{q=p}^{p+k-1}\frac{\bar{G}_{q} (\bar{X}_{q}
)}{A_{q+1}^{\mathbb{I}_{\bar{C}_{d}^{c}} (\bar{X}_{q}
)}B_{q+1}^{\mathbb{I}_{\bar{C}_{d}} (\bar{X}_{q} )}}
\Biggr]\bar{v} (\bar{X}_{p+k} ),\qquad 1\leq k\leq n-p.
\end{eqnarray*}
Then for $1\leq k\leq n-p$, $\bar{\mathbb{E}}_{p+k-1,\bar
{X}_{p+k-1}} [\Xi_{k} ]\leq\Xi_{k-1}$,
so that
%
%e63 #&#
\begin{equation}\label{eqfilsupermart}
\bar{\mathbb{E}}_{p,\delta_{x}\otimes\eta_{p}} [\Xi_{n-p} ] \leq \bar{
\mathbb{E}}_{p,\delta_{x}\otimes\eta_{p}} [\Xi_{0} ]=v(x)\eta_{p} (v )
\leq cv(x)\mu(v ),
\end{equation}
where the last inequality is due to Proposition \ref{propetaboundedinv}.

By Douc et~al. [(\citeyear{filtertheDFMP09}), Lemma 19], $\bar{M}_{p,n}^{ (d
)}<\beta(n-p )$
implies $\widetilde{M}{}^{ (d )}_{p,n}\geq a_{p,n}$, and then
%
%e64 #&#
\begin{eqnarray}\label{eqfilABbound}
& & \Biggl[\prod_{q=p}^{p+k-1}A_{q+1}^{\mathbb{I}_{\bar{C}_{d}^{c}}
(\bar{X}_{q} )}B_{q+1}^{\mathbb{I}_{\bar{C}_{d}} (\bar
{X}_{q} )}
\Biggr]\mathbb{I} \bigl\{ \bar{M}_{p,n}^{ (d
)}<\beta(n-p )
\bigr\}
\nonumber
\\
& &\qquad \leq\Bigl(\sup_{q\geq1}A_{q}
\Bigr)^{a_{p,n}} \biggl(1\vee\sup_{q\geq
1}\interleave
Q_{q}\interleave_{v} \biggr)^{2 (n-p-a_{p,n}
)}
\nonumber
\\
& &\qquad \leq\biggl(\sup_{q\geq1}\interleave\mathbb
{I}_{C_{d}^{c}}Q_{q}\interleave_{v}
\biggr)^{a_{p,n}} \biggl(1\vee\sup_{q\geq1}\interleave
Q_{q}\interleave_{v} \biggr)^{2 (n-p
)}
\\
& &\qquad \leq e^{-\delta da_{p,n}} \biggl(1\vee\sup_{q\geq1}\interleave
Q_{q}\interleave_{v} \biggr)^{2 (n-p )}
\nonumber
\\
& &\qquad \leq\exp(-\delta da_{p,n} )\exp\bigl[0\vee2b_{\underline
{d}}
(n-p ) \bigr],\nonumber
\end{eqnarray}
where \ref{hypdrift} has been used. For the remainder of the
proof we may assume without loss of generality that $b_{\underline{d}}>0$.

Combining (\ref{eqfilsupermart}) and (\ref{eqfilABbound})
then gives
\begin{eqnarray*}
\Delta_{p,n}^{(2)} (x,\eta_{p} ) & \leq& \bar{
\mathbb{E}}_{\delta_{x}\otimes\eta_{p}} \Biggl[\prod_{q=p}^{n-1}
\bar{G}_{q} (\bar{X}_{q} )\mathbb{I} \bigl\{
\bar{M}_{p,n}^{ (d
)}<\beta(n-p ) \bigr\} \bar{v} (
\bar{X}_{n} ) \Biggr]
\\
% & \leq& \left(\sup_{q\geq1}A_{q}\right)^{a_{p,n}}\left(\sup_{q
& \leq& c\exp\bigl[-\delta
da_{p,n}+2b_{\underline{d}} (n-p ) \bigr]v(x)\mu(v )
\end{eqnarray*}
and therefore
%
%e65 #&#
\begin{eqnarray}\label{eqDelta2bound}
& & \frac{\Delta_{p,n}^{(2)} (x,\eta_{p} )}{Q_{p,n}
(1 )(x)\eta_{p}Q_{p,n} (1 )}
\nonumber
\\
& &\qquad =\frac{\Delta_{p,n}^{(2)} (x,\eta_{p} )}{h_{p,n}(x)
(\prod_{q=p}^{n-1}\lambda_{q} )^{2}}
\\
& &\qquad \leq c\exp\bigl[-\delta da_{p,n}+2 (n-p ) (b_{\underline{d}}-\log
\underline{\lambda} ) \bigr]\frac
{v_{p,n}(x)}{\llVert h_{p,n}\rrVert_{v}}\mu(v ),\nonumber
\end{eqnarray}
where Propositions \ref{propetaboundedinv} and \ref{propequiv}
have been applied and
$\underline{\lambda}=\inf_{n\geq0}\lambda_{n}>0$.\eject

Collecting the bounds of (\ref{eqDelta1bound}), (\ref{eqDelta2bound})
and returning to (\ref{eqfiltstartingdecomp}), we establish that
\begin{eqnarray*}
& & \biggl\llvert\frac{Q_{p,n} (\varphi)(x)}{h_{p,n}(x)\prod
_{q=p}^{n-1}\lambda_{q}}-\eta_{n} (\varphi)\biggr
\rrvert
\\
& &\qquad \leq2c\llVert\varphi\rrVert_{v}\frac{v_{p,n}(x)}{\llVert
h_{p,n}\rrVert_{v}}\mu(v )
\bigl[\rho_{d}^{\beta
(n-p )}+\exp\bigl[-\delta da_{p,n}+2
(n-p ) (b_{\underline{d}}-\log\underline{\lambda} ) \bigr] \bigr]
\\
& &\qquad \leq2c\llVert\varphi\rrVert_{v}\frac{v(x)}{h_{p,n}(x)}\mu(v )
\\
& &\qquad\quad{} \times\bigl[\rho_{d}^{\beta(n-p )}+\exp\bigl[- (n-p ) \bigl(
\delta d(1-\beta)/2-2b_{\underline{d}}+2\log\underline{\lambda} \bigr
)+3\delta
d/2 \bigr] \bigr],
\end{eqnarray*}
where for the second inequality, $ \lfloor a \rfloor\geq a-1$
has been used. The proof is complete upon recalling that $d\in
[\underline{d},\infty)$
was arbitrary, $\rho_{d}<1$, $\beta\in(0,1)$ and multiplying through
by $h_{p,n}(x)$.
\end{pf*}

\subsection*{Proofs for Section \protect\ref{subExponential-moments-for}}

\mbox{}

\begin{pf*}{Proof of Theorem \ref{thmexpmoments}}
Throughout, the proof $c$ denotes a finite constant whose value may
change on each appearance, but which depends only on $\mu$ and the
quantities in \ref{hypdrift}--\ref{hypminorandmaj}. Also,
throughout the proof we take by convention that for any $j<k$,
$\sum_{k}^{j}\equiv0$.

First consider the case $s>0$. By Lemma \ref{lemchangeofmeasure},
%
%e66 #&#
\begin{eqnarray}\label{eqexpmomentstwisted}
& & \frac{\mathbb{E}_{\mu} [\prod_{p=0}^{n-1}G_{p} (X_{p}
)\exp(\sum_{k\in\{ i_{1},\ldots,i_{s} \} }\llvert F_{k} (X_{k}
)\rrvert ) ]}{\mathbb{E}_{\mu}
[\prod_{p=0}^{n-1}G_{p} (X_{p} ) ]}
\nonumber
\\
& &\qquad =\int\mu(dx )h_{0,n} (x )\check{\mathbb{E}}_{x}^{ (n )}
\biggl[\exp\biggl(\sum_{k\in\{ i_{1},\ldots,i_{s} \} }\bigl\llvert
F_{k} (\check{X}_{k,n} )\bigr\rrvert\biggr) \biggr]
\\
& &\qquad \leq\biggl(\prod_{k\in\{ i_{1},\ldots,i_{s} \} }\bigl\llVert
e^{\llvert F_{k}\rrvert}\bigr\rrVert_{v^{\delta}} \biggr)\int\mu
(dx)h_{0,n}(x)
\check{\mathbb{E}}_{x}^{ (n )} \biggl[\prod
_{k\in
\{ i_{1},\ldots,i_{s} \} }v^{\delta} (\check{X}_{k,n} )
\biggr].\nonumber
\end{eqnarray}
We now obtain some bounds which will be used to control the expectation
in (\ref{eqexpmomentstwisted}). Proposition \ref{propetaboundedinv}
holds under the assumptions of the theorem so we may apply the upper
and lower bounds of Proposition \ref{propequiv} and Lemma \ref
{lemhpnboundedbelowonC}
to the bound of Lemma~\ref{lemdriftbound} and choose $d$ therein
large enough, in order to establish that there exists a finite constant
$c$ independent of $1\leq p<q\leq n$ and $x\in\mathsf{X}$ such
that
%
%e67 #&#
\begin{eqnarray}\label{eqtwisteddriftboundc1}
& & \check{\mathbb{E}}_{p,x}^{ (n )} \bigl[v_{q,n} (
\check{X}_{q,n} ) \bigr]
\nonumber
\\
& &\qquad \leq\frac{e^{-\delta d (q-p )}}{\underline{\lambda}^{
(q-p )}}\frac{\llVert h_{q,n}\rrVert_{v}}{\llVert
h_{p,n}\rrVert_{v}}v_{p,n} (x )
\nonumber\\[-8pt]\\[-8pt]
& &\qquad\quad{} +\frac{e^{d(1-\delta)+b_{d}}}{\varepsilon_{d}^{-}}\llVert
h_{q,n}\rrVert_{v} \Biggl[
\frac{1}{\nu_{d} (\mathbb
{I}_{C_{d}}h_{q,n} )}+\sum_{k=p+1}^{q-1}
\frac{e^{-\delta d
(q-k )}}{\underline{\lambda}^{ (q-k )}}\frac{1}{\nu
_{d} (\mathbb{I}_{C_{d}}h_{k,n} )} \Biggr]
\nonumber
\\
&&\qquad \leq c\frac{\llVert h_{q,n}\rrVert_{v}}{\llVert
h_{p,n}\rrVert_{v}}v_{p,n} (x ),\nonumber
\end{eqnarray}
where $\underline{\lambda}=\inf_{n\geq0}\lambda_{n}>0$. Therefore
by \ref{hypdrift}, for $p\leq q$,
%
%e68 #&#
\begin{eqnarray}\label{eqtwisterdriftboundc2}
v^{\delta}(x)\check{\mathbb{E}}_{p-1,x}^{ (n )}
\bigl[v_{q,n} (\check{X}_{q,n} ) \bigr] & \leq&
cv^{\delta}(x)\frac
{\llVert h_{q,n}\rrVert_{v}}{\llVert h_{p,n}\rrVert
_{v}}S_{p,n} (v_{p,n} ) (x
)
\nonumber
\\
& = & cv^{\delta}(x)\llVert h_{q,n}\rrVert_{v}
\frac{Q_{p}
(v ) (x )}{\lambda_{p-1}h_{p-1,n}(x)}
\nonumber\\[-8pt]\\[-8pt]
& \leq& c\llVert h_{q,n}\rrVert_{v}\frac{v (x
)}{h_{p-1,n}(x)}
\frac{e^{b_{\underline{d}}}}{\underline{\lambda
}}
\nonumber
\\
& \leq& c\frac{\llVert h_{q,n}\rrVert_{v}}{\llVert
h_{p-1,n}\rrVert_{v}}v_{p-1,n}(x).\nonumber
\end{eqnarray}
Now fix $n\geq1$, $1\leq s\leq n+1$, $ (i_{1},\ldots,i_{s}
)\in\mathcal{I}_{n,s}$
arbitrarily and define $ (\Xi_{k,n};0\leq k\leq s )$ by
\begin{eqnarray*}
\Xi_{0,n}&:=&\frac{v_{0,n} (\check{X}_{0,n} )}{\llVert
h_{0,n}\rrVert_{v}},
\\
\Xi_{k,n}&:=&\frac{v_{i_{k},n} (\check{X}_{i_{k},n} )}{\llVert
h_{i_{k},n}\rrVert_{v}}\exp\Biggl[\sum
_{j=1}^{k-1} \bigl(\delta V (\check{X}_{i_{j},n}
)-\log c \bigr) \Biggr],\qquad 1\leq k\leq s,
\end{eqnarray*}
where $c$ is as in (\ref{eqtwisterdriftboundc2}). We then have
\begin{eqnarray*}
& & \check{\mathbb{E}}_{i_{k-1},\check{X}_{i_{k-1},n}}^{ (n
)} [\Xi_{k,n} ]
\\
& &\qquad =\frac{1}{\llVert h_{i_{k},n}\rrVert_{v}}\check{\mathbb
{E}}_{i_{k-1},\check{X}_{i_{k-1},n}}^{ (n )}
\bigl[v_{i_{k},n} (\check{X}_{i_{k},n} ) \bigr]\exp\Biggl[\sum
_{j=1}^{k-1} \bigl(\delta V (
\check{X}_{i_{j},n} )-\log c \bigr) \Biggr]
\\
& &\qquad \leq c\frac{v_{i_{k-1},n} (\check{X}_{i_{k-1},n} )}{\llVert
h_{i_{k-1},n}\rrVert_{v}}v^{\delta} (\check{X}_{i_{k-1},n} )\exp
\Biggl[\sum_{j=1}^{k-1} \bigl(\delta V (
\check{X}_{i_{j},n} )-\log c \bigr) \Biggr]
\\
& &\qquad =\Xi_{k-1,n},
\end{eqnarray*}
where the inequality is due to (\ref{eqtwisterdriftboundc2}).
Thus $ (\Xi_{k,n},\check{\mathcal{F}}_{k,n};0\leq k\leq s )$
is a super-Martingale, with $\check{\mathcal{F}}_{k,n}:=\sigma
(\check{X}_{0,n},\ldots,\check{X}_{i_{k}-1,n},\check{X}_{i_{k},n} )$.
Therefore
%
%e69 #&#
\begin{eqnarray}\label{eqtwistedsupermart}
\check{\mathbb{E}}_{x}^{ (n )} \biggl[\prod
_{k\in\{
i_{1},\ldots,i_{s} \} }v^{\delta} (\check{X}_{k,n} ) \biggr] &
\leq& c^{s}\check{\mathbb{E}}_{x}^{ (n )} [\Xi
_{s,n} ]\llVert h_{i_{s},n}\rrVert_{v}
\nonumber
\\
& \leq& c^{s}\frac{v_{0,n} (x )}{\llVert h_{0,n}\rrVert_{v}}\llVert
h_{i_{k},n}\rrVert
_{v}
\\
& \leq& c^{s}\frac{v (x )}{h_{0,n}(x)},\nonumber
\end{eqnarray}
where Propositions \ref{propetaboundedinv} and \ref{propequiv}
have been used for the last inequality.

The proof is completed upon combining (\ref{eqtwistedsupermart})
with (\ref{eqexpmomentstwisted}) and noting that the result holds
trivially when $s=0$.
\end{pf*}


%suskaldyti doi

% imsref loaded by lrinkeviciute, 2013-05-10 15:57:34

\printaddresses

\end{document}